\documentclass[pra,twocolumn]{revtex4}
\usepackage{amsmath}
\usepackage[dvips]{graphicx}
\usepackage{layout}
\usepackage{float}
\begin{document}

\title{Bose-Fermi Pair Correlations in Attractively Interacting Bose-Fermi Atomic 
Mixtures}
\author{Takayuki Watanabe$^1$} 
\author{Toru Suzuki$^1$} 
\author{Peter Schuck$^{2,3,4}$}
\affiliation{$^1$ Department of Physics, Tokyo Metropolitan University,  Hachioji, 
Tokyo 192-0397, Japan}
\affiliation{$^2$ Institut de Physique Nucl\'eaire, IN2P3-CNRS, UMR8608, F-91406 Orsay, France}
\affiliation{$^3$ Universit\'e Paris-Sud, F-91406 Orsay, France}
\affiliation{$^4$ Laboratoire de Physique et Mod\'elisation des Milieux Condens\'es, CNRS \& Universit\'e Joseph Fourier, 
Maison des Magist\`eres, Bo\^ite Postale 166, 38042 Grenoble Cedex 9, France}
\date{\today}

\begin{abstract}
        We study  static properties of attractively interacting Bose-Fermi mixtures 
        of uniform atomic gases  at zero temperature.  Using Green's function formalism 
         we calculate boson-fermion scattering amplitude     and fermion self-energy in 
        the medium to lowest order of the hole line expansion. 
        We study ground state energy and pressure as functions of the scattering length 
        for a few values of the boson-fermion mass ratio $m_b/m_f$ and the number 
        ratio $N_b/N_f$. We find that the attractive contribution to energy is greatly enhanced for 
        small values of the mass ratio. 
        We study the role of the Bose-Fermi pair correlations in the mixture 
        by calculating the pole of the boson-fermion scattering amplitude in the 
        medium. The pole shows a standard quasiparticle dispersion for 
        a Bose-Fermi pair, for $m_b/m_f\geq 1$. For small values of the mass ratio, 
        on the other hand, a Bose-Fermi pair with a finite center-of-mass momentum 
        experiences a strong attraction, implying large medium effects.
	  In addition, we also study the fermion dispersion relation. 
	  We find two dispersion branches with the possibility of the avoided crossings. 
	  This strongly depends on the number rario $N_b/N_f$.
\end{abstract}
\maketitle

\def\be{\begin{equation}}
\def\ee{\end{equation}}
\def\intdx{\int\!d^3{\bf x}}
\def\xbf{{\bf x}}
\def\psix{\psi({\bf x})}
\def\psidx{\psi^\dagger({\bf x})}
\def\vpx{\varphi({\bf x})}
\def\vpdx{\varphi^\dagger({\bf x})}
\def\half{\frac{1}{2}}
\def\kbf{{\bf k}}
\def\intdq{\int\! \frac{d^4q}{(2\pi)^4}}
\def\intdk{\int\! \frac{d^3{\bf k}}{(2\pi)^3}}

\section{Introduction}

Recent developments in the  field of cold atomic gases have proven that this 
system provides an ideal laboratory for the studies of quantum many-body 
systems\cite{pethick}.
This is due to the experimental facilities which allow 
to control various parameters characterizing the many-body system, e.g., external 
potentials including optical lattices, choice of atoms obeying Bose or Fermi statistics
and their mixtures, variable particle densities, etc. The use of Feshbach resonances, 
in particular, makes the atomic gases 
an extremely flexible system as it provides a means to control atomic 
interactions\cite{fesh_th,fesh_exp}. One thus was, for instance, able to study the  
BEC-BCS crossover process in the two-component Fermi system, which has 
been under intense investigation for decades\cite{crossover}. 
By changing the resonance energies through the external magnetic field, 
one can in principle change the magnitude and the sign of the scattering length 
of the interacting particles, keeping track all the way from a resonating fermion pair to a bound 
composite particle, a bosonic molecule. 

The aim of the present paper is the study of pair correlations in a different system, 
a Bose-Fermi(BF) mixture. Degenerate mixtures of bosons and fermions have been created 
since several years, and studies of static and dynamic properties have been 
performed\cite{BFexp}. Among those are the studies of attractively interacting BF systems, 
where one finds a sudden loss of fermions as the BF attractive interaction is effectively 
increased\cite{BFcollapse}. Detailed studies of the dynamics of this system are still 
missing, however. Recently the finding of  Feshbach resonances and formation of the 
boson-fermion molecules have been reported\cite{BFres,BFmol}. It is thus expected that 
by controling the BF interaction one may realize an analog of the process found in 
two-component Fermi systems. What should be expected if one replaced fermion pairs in 
the BEC-BCS crossover process by BF pairs? Such studies have indeed been 
performed theoretically\cite{schuck} (see also \cite{bf_pair}.) By adopting a 
Cooper type two-particle problem on top of the boson-fermion degenerate system, 
it was shown  that a stable correlated BF pair is created 
even before the threshold for the BF bound state. In contrast to the BCS case, 
however, the system allows only one correlated BF pair with a given center-of-mass (CM)
momentum because of the fermionic nature of the composite particle. It is then 
suggested that by increasing the BF attractive interaction, 
one may create BF pairs with different CM momentum stepwise, until finally a new 
Fermi sea of the BF pairs is completed. 

In Ref.\cite{schuck} a separable BF interaction has been adopted to elucidate 
the mechanism of the creation of BF pairs. In the present paper we adopt a 
standard pseudopotential for the interaction, and calculate energy and pressure 
of the system for various values of input parameters. We use Green's function 
formalism for this system and calculate perturbatively relevant diagrams to 
lowest order of the hole-line expansion. Such formalism has been developed 
in \cite{albus} together with the calculation of the energies including 
Bose-Bose (BB) interaction. Our formulation is similar to \cite{albus},  
but we use the renormalization procedure of \cite{randeria} in relating the 
pseudopotential strength to the $S$-wave scattering length. This allows us to 
formally take the limit $|a|\rightarrow\infty$, the unitarity limit
\cite{unit_th}, which 
is necessary when one considers a (nearly) bound state of a pair of atoms. 
We then calculate  the poles of the BF pair scattering amplitude in 
the BF medium, which may be compared with the results of \cite{schuck}. 
Studies of the behavior of the poles as a function of input parameters 
give us suggestions on the role of the BF pair correlations in the 
static properties of the system.

The content of the paper is as follows: In the next section we present our model 
based on the Hamiltonian without Bose-Bose interaction. We calculate the BF 
scattering amplitude in the BF mixture in ladder approximation, and give formulas 
for physical quantities in terms of the amplitude.  In section 3 we show  numerical 
results for the ground 
state energy and pressure for various choices of the boson/fermion masses and the 
values of the Bose-Fermi interaction. We then study  Bose-Fermi pair correlation in 
Section 4 by focusing on the pole structure of the boson-fermion scattering amplitude 
in the mixture. 
We also calculate the pole of the single fermion Green's function 
and study the role of the Bose-Fermi pair and its dispersion.
We summarize our results in section 5 together with a comment on the 
effects of the Bose-Bose interaction. 
Detailed expressions for the scattering amplitude are given in the appendix. 

\section{Formulation}

We consider a uniform system of a polarized Bose-Fermi mixture of atomic gases
with attractive boson-fermion interaction. The model Hamiltonian of the system is
given by
\begin{align}
       H & = T_b+T_f+H_{bf}, \nonumber \\
     T_b & = \intdx \phi^{\dagger} ({\bf x}) 
            \left( -\frac{\nabla^2}{2m_b} - \mu_b \right) \phi({\bf x}), \nonumber \\
     T_f & = \intdx \psidx \left( -\frac{\nabla^2}{2m_f} \right) \psix, \nonumber \\
     H_{bf} & =  g_{bf} \intdx \phi^{\dagger} ({\bf x}) \psidx \psix \phi({\bf x}),
     \end{align}
where $\psi$ and $\phi$ are the boson and fermion field operators, respectively,
$T_b$ and $T_f$ denote bosonic and fermionic kinetic energies, while $H_{bf}$
denotes boson-fermion interaction with  strength $g_{bf}\,(<0)$ of the boson-fermion
pseudopotential. Effects of the boson-boson interaction will be mentioned later, while 
the fermion-fermion interaction is omitted throughout as we consider 
one-component (polarized) fermions. We will adopt the Bogoliubov approximation in treating 
the Bose-Einstein condensate (BEC), and therefore include in $T_b$ the bosonic
chemical potential $\mu_b$.

\subsection{Green's function formalism in the Bose-Fermi mixture}

To treat condensed bosons, we adopt the conventional Bogoliubov method by separating 
the zero momentum mode from the remainder :
\begin{equation}
        \phi({\bf x}) = \sqrt{n_0} + \vpx \label{eq:shiftope}
        \end{equation}
together with its conjugate. $n_0=N_0/V$ is the number density of bosons with momentum 
${\bf k}=0$. As usual we omit the fluctuation of the boson number in the condensate. 
The boson number operator $\hat{N}_b$ writes
\begin{equation}
        \hat{N}_b = N_0 + \int d^3{\bf x} \vpdx \vpx.
        \end{equation}
and the Hamiltonian takes the form
\begin{equation}
        H = H_0 + H_{bf},
        \end{equation}
where
\begin{align}
      H_0 & = \intdx \vpdx \left(-\frac{\nabla^2}{2m_b}-\mu_b \right)\vpx \nonumber \\
          & + \intdx \psidx \left( -\frac{\nabla^2}{2m_f} \right) \psix \nonumber \\
          &  - \mu_b N_0
       \label{eq:freeHami}
       \end{align}
and
\begin{align}
        H_{bf} & =  n_0 g_{bf} \intdx \psidx \psix \nonumber \\
               & + \sqrt{n_0} g_{bf} \intdx \psidx \psix (\vpdx+\vpx ) \nonumber \\
              & + g_{bf} \intdx \psidx \vpdx \vpx \psix.
        \label{eq:intHami}
        \end{align}
Physical quantities  can be expressed in terms of Green's functions. We define the 
boson and fermion Green's functions by
\begin{align}
     iG^f(x-y) & = \frac{\langle \Psi_0 | T \left[ \psi_H (x) \psi^{\dagger}_H(y) 
                    \right] | \Psi_0 \rangle}  {\langle \Psi_0 | \Psi_0 \rangle},
        \label{eq:fpro} \\
     iG^b(x-y) & = \frac{\langle \Psi_0 | T \left[ \varphi_H (x) \varphi^{\dagger}_H(y)
                      \right] | \Psi_0 \rangle}  {\langle \Psi_0 | \Psi_0 \rangle},
      \label{eq:bpro} 
      \end{align}
where $\psi_H(x), \varphi_H(x)$ are the field operators in the Heisenberg picture, 
and $| \Psi_0 \rangle$ represents the interacting ground state.

The energy of the system can be expressed in terms of the Green's functions. The fermion 
and boson kinetic energies are calculated according to the standard procedure 
\cite{fetter} as:
\begin{align}
        \langle T_f \rangle  & = \left< \frac{-\nabla^2}{2m_f} \right>
                 = -i V \intdq \epsilon^{f}_{\bf q} G^f(q) e^{iq_0 \eta} ,\\
        \langle T_b \rangle & = \left< \frac{-\nabla^2}{2m_b} \right>
                 = i V \intdq \epsilon^{b}_{\bf q} G^b(q) e^{iq_0 \eta},
        \end{align}
where $G(q)$'s are the Fourier transform of the Green's functions, $\eta$ is a positive 
infinitesimal, and we set $\epsilon^{b,f}_{\bf q}={\bf q}^2/2m_{b,f}$. The different 
signs in the two expressions come from the ordering of the field operators.

To calculate the interaction energy, we first consider  the Heisenberg equation of 
motion for the fermion field:
\begin{align}
        i \frac{\partial}{\partial t}\psi_H (x) = &  [ \psi_H(x), H ] \nonumber \\
         = & ( -\frac{\nabla^2}{2m_f} ) \psi_H(x) + n_0 g_{bf} \psi_H(x) \nonumber \\
           & + \sqrt{n_0} g_{bf}\psi_H(x) ( \varphi_H^\dagger(x)+ \varphi_H(x)) \nonumber \\
               & +  g_{bf} \psi_H(x)  \varphi^{\dagger}_H(x)  \varphi_H(x) .
        \end{align}
Multiplying by $\psi^{\dagger}_H(x')$  and integrating over ${\bf x}$, we obtain
\begin{align}
        \left< H_{bf} \right> & = -i \intdx  \lim_{\begin{smallmatrix} {\bf x}' \to {\bf x} \\ t' \to t \end{smallmatrix}}
                   \left( i\frac{\partial}{\partial t} + \frac{\nabla^2}{2m_f} \right)
                        G^f(t{\bf x}, t'{\bf x}') \nonumber \\
                & = -i V \intdq ( q_0 - \epsilon^{f}_{\bf q} ) G^f(q) e^{iq_0 \eta} .
        \label{eq:vbf_temp}
        \end{align}
We now introduce fermion and boson self-energies $\Sigma^f(q)$ and $\Sigma^b(q)$ through
\begin{equation}   
          G^f(q) =  \frac{1}{q_0 - \epsilon^{f}_{\bf q}-\Sigma^{f}(q)} ,
          \label{eq:Gfermi}
       \end{equation}
and
\begin{equation}   
        G^b(q) = \frac{1}{q_0 - \epsilon^{b}_{\bf q} + \mu_b - \Sigma^{b}(q)}.
        \label{eq:Gbose}
       \end{equation}
In the integrand of Eq.(\ref{eq:vbf_temp})  one may use the relation from 
Eq.(\ref{eq:Gfermi})
\begin{equation}   
         \left( q_0 - \epsilon^{f}_{\bf q} \right) G^f(q)=1+\Sigma^f(q)G^f(q),
         \label{eq:Gfeq}
       \end{equation}
and finds
\begin{equation}   
        \langle H_{bf} \rangle = -i V \intdq \Sigma^f(q) G^f(q) e^{iq_0 \eta} ,
       \end{equation}
the first term in the r.h.s. in Eq.(\ref{eq:Gfeq}) giving a null contribution 
to the integral. The total energy $E$ of the system is finally obtained as
\begin{align}
    E = & \langle T_f \rangle + \langle T_b \rangle + \langle H_{bf} \rangle 
                                                                     \nonumber \\
      = & -iV \intdq \left(\epsilon_{\bf q}^f + \Sigma^f(q) \right) 
                                                  G^f(q) e^{iq_0 \eta} \nonumber \\
        & +iV \intdq \epsilon^b_{\bf q} G^b(q) e^{iq_0 \eta}.
         \label{eq:ene_exp} 
        \end{align}
The thermodynamic potential at zero temperature is given by
\begin{equation}
        \Omega(N_f, N_0, \mu_b) = \langle {H} \rangle
                                =  E - \mu_b \langle \hat{N}_b \rangle,
        \end{equation}
where
\begin{equation}
        \langle \hat{N}_b \rangle = N_0 + iV \intdq G^b(q) e^{iq_0 \eta}.
        \end{equation}

The system is characterized by the boson and fermion particle numbers, $N_b$ 
and $N_f$. The number of bosons satisfies the thermodynamic relation 
\begin{equation}
        \frac{\partial \Omega}{\partial \mu_b}  = - N_b.
	  \label{eq:therm}
        \end{equation}
The parameter  $N_0$ should be chosen to minimize the thermodynamic potential
\begin{equation}
        \frac{\partial \Omega}{\partial N_0}  = 0
        \label{eq:bnum}
        \end{equation}
which leads to an explicit expression for $\mu_b$ as shown below.

We also will calculate the pressure to discuss the stability of the system: As usual, it
is obtained from  the thermodynamic relation
\begin{equation}
        P = \frac{\partial E}{\partial V} .
       \end{equation}

\subsection{Self-energy in the ladder approximation}

\begin{figure}[htb]
        \begin{center}
        \includegraphics[scale=.7]{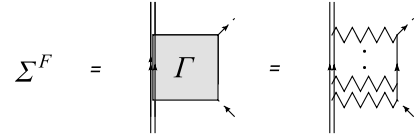}
        \end{center}
        \caption{Fermion self-energy in ladder approximation.  The double solid 
        lines  represent fermion propagation, while single solid line represents 
        noncondensed free  boson propagation.  The arrows denote  condensed bosons 
        and are associated with the  factor $\sqrt{n_0}$. The zigzag lines represent the 
        boson-fermion interaction $g_{bf}$. }
        \label{fig:selfenergy}
\end{figure}

\begin{figure}[htb]
        \begin{center}
        \includegraphics[scale=0.5]{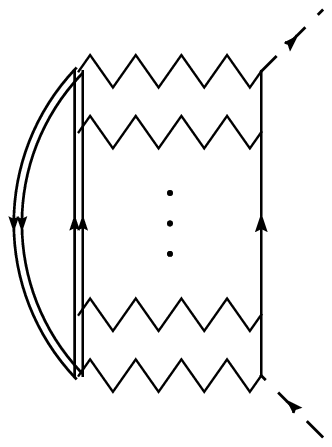}
        \end{center}
	  \caption{Ladder diagram contribution to the interaction energy. The downward 
        double  solid line indicates a hole propagation. Otherwise as in fig.\ref{fig:selfenergy}.}
	  \label{fig:enedi}
\end{figure}

To obtain the total energy of the system we  calculate the fermion self-energy 
$\Sigma^F$ in ladder approximation. Here the self-energy is expressed in terms 
of the two-particle scattering amplitude, 
$\Gamma({\bf q},{\bf q'},P)$, in the medium of the Bose-Fermi mixture as shown in 
fig.\ref{fig:selfenergy}. The interaction energy is accordingly calculated up to 
the lowest two-particle correlation diagram, fig.\ref{fig:enedi},   
in the spirit of the hole-line expansion \cite{fetter}.


The scattering amplitude $\Gamma$ in the present model obeys the integral equation 
\cite{albus,schuck,fetter,galli},
\begin{widetext}
\begin{align}
        \Gamma({\bf q}, {\bf q'}, P) = g_{bf} + ig_{bf} \int \frac{d^4k}{(2\pi)^4}
            G^f_{0} \left( \frac{m_f}{m_f+m_b} P + k \right) G^b_{0} \left( \frac{m_b}{m_f+m_b} P - k \right)
             \Gamma({\bf k},{\bf q'},P) ,
        \label{eq:gallim}
\end{align}
\end{widetext}
where $P$ denotes a four-momentum of the center-of-mass motion of the interacting 
particles, while ${\bf q}$ and ${\bf q'}$ are the relative three-momentum in the final 
and intial states.  The boson and fermion free Green's functions in medium are given by
\begin{align}
        G^f_0(p) & = \frac{ \theta \left( |{\bf p}| - k_F \right) }{ p_0 - \epsilon_{\bf p}^f 
                   + i\eta}  + \frac{ \theta \left( k_F - |{\bf p}| \right) }{ p_0 - 
                   \epsilon_{\bf p}^f - i\eta} , \displaybreak[0]\\
        G^b_0(p) & = \frac{ 1 }{ p_0 - \epsilon_{\bf p}^b + \mu_b + i\eta} ,
        \label{eq:freeGF}
\end{align}
where the Fermi momentum $k_F$ is fixed by the fermion density $N_f/V$. After the 
integration over $k_0$, Eq.(\ref{eq:gallim}) becomes
\begin{widetext}
\begin{align}
        \Gamma( {\bf q},  {\bf q'}, P) = g_{bf} + g_{bf} \intdk
         \frac{ \theta \left( \left| \tilde{{\bf P}}_f + {\bf k} \right| - k_F \right) }
        {P_0-\epsilon^f_{{\bf \tilde{P}}_f + {\bf k}} -\epsilon^b_{{\bf \tilde{P}}_b -{\bf k}}
               + \mu_b +i \eta}  \Gamma({\bf k},{\bf q'},P)
        \label{eq:temp1}
\end{align}
\end{widetext}
with ${\bf \tilde{P}}_f = m_f/(m_f+m_b){\bf P}$ and ${\bf \tilde{P}}_b = m_b/(m_f+m_b){\bf P}$. 
We dropped the hole propagation part in accordance with the present approximation. 
With respect to the T-matrix equation in \cite{schuck}, we notice that there the phase space factor in (\ref{eq:temp1})
is replaced by 
$
 \theta \left( \left| \tilde{\bf P}_f + {\bf k} \right| - k_F \right) 
\rightarrow 
  \theta \left( \left| \tilde{\bf P}_f + {\bf k} \right| - k_F \right) + N_0
$.
This is natural, because in \cite{schuck} the shift operation (\ref{eq:shiftope}) for the bosons has
not been performed and therefore the free boson occupancy $N_0$ appears additionally. 
The two formulations are, however, essentially equivalent.
From the structure of Eq.(\ref{eq:temp1}), one easily finds that $\Gamma$ depends only on the 
variable $P$, and we hereafter write simply $\Gamma(P)$. One also finds that the 
integral in Eq.(\ref{eq:temp1}) requires a momentum cutoff, which originates from
the use of the zero-range interaction. We can remedy this shortcoming by employing the 
observable S-wave scattering length $a$, instead of the pseudopotential 
coupling constant $g_{bf}$.
We perform this renormalization following the procedure adopted in \cite{randeria} (see 
also, \cite{ohashi}), 
slightly different from the one in \cite{albus}. The S-wave scattering length is 
related to the two-particle scattering amplitude $\Gamma_0$ in vacuum by the relation
\begin{equation}
        \Gamma_0({\bf q}={\bf q'}=P=0)=\frac{2\pi a}{\nu},
       \end{equation}
where $\nu$ is the reduced mass, and $\Gamma_0$ obeys the equation similar to 
Eq.(\ref{eq:gallim}) with $G_0$ replaced with the free Green's function in vacuum. 
By solving the equation for $\Gamma_0$ one obtains
\begin{equation}
       \frac{2\pi a}{\nu} =  \frac{g_{bf}}{ \displaystyle{ 1+g_{bf}\intdk
                    \frac{1}{\epsilon^f_{\bf k}+\epsilon^b_{\bf k} } } }
       \label{eq:renorm},
       \end{equation}
where the integral in the denominator involves again the implicit momentum cutoff.
Now one may combine the above expression with Eq.(\ref{eq:temp1}), and eliminate 
$g_{bf}$ in favor of the scattering length $a$, and finally obtains
\begin{align}
       \Gamma(P)=\frac{2\pi a}{\nu} \Bigg[ 1- \frac{2\pi a}{\nu} I\left(P_0, \left|{\bf P}\right| \right) \Bigg]^{-1} 
        \label{eq:gammap}
\end{align}
with
\begin{widetext}
\begin{equation}
        I\left( P_0, \left|{\bf P}\right| \right) = \int \frac{d^3{\bf k}}{\left( 2\pi \right)^3}
         \left\{ \frac{ \theta \left( \left| \tilde{\bf P}_f + {\bf k} \right| - k_F \right) }
         {P_0-\epsilon^f_{{\bf \tilde{P}}_f + {\bf k}} -\epsilon^b_{{\bf \tilde{P}}_b -{\bf k}} + \mu_b +i \eta}
         + \frac{1}{\epsilon^f_{\bf k} + \epsilon^b_{\bf k} } \right\}.
\end{equation}
\end{widetext}
Since the integral in the denominator is convergent at large $| {\bf k} |$, we can 
let the momentum cutoff  go to infinity.  The expression (\ref{eq:gammap}) involves 
all orders in the scattering length and allows us to formally take the unitarity 
limit $|a| \rightarrow \infty$ in the following section. This limit has been studied 
for two-component Fermi systems in relation with the BEC to BCS crossover phenomenon. 
If a similar phenomenon is expected or not for 
Bose-Fermi pairs will be studied in the next section.  

Using above vertex function, we can calculate the proper self-energies for the fermion 
and the boson as
\begin{align}
        \Sigma^f (p) & = n_0 \Gamma ( p ) \label{eq:fself}\\
        \Sigma^b (p) & = -i \int \frac{d^4 p'}{( 2\pi)^4} G^f_0 (p') \Gamma(p+p')
        \label{eq:bstemp} .
        \end{align}
Expression (\ref{eq:temp1}) implies that $\Gamma(p+p')$ is analytic in the upper 
half $p_0'$ plane, and Eq.(\ref{eq:bstemp}) reduces to
\begin{equation}
        \Sigma^b (p) = \int \frac{d^3 {\bf p'}}{( 2\pi )^3} \theta 
                 ( k_F - | {\bf p'} | )  \Gamma(p + p').
         \label{eq:bself}
        \end{equation}
with $p_0'=\epsilon^f_{\bf p'}$. This shows that $\Sigma^b(p)$, and  hence, also $G^b(p)$ 
is  analytic in the upper half $p_0$ plane. One then finds from Eq.(\ref{eq:therm}) 
that
\begin{equation}
        N_b = N_0 
        \label{eq:connum} .
        \end{equation}

\section{Results for Total Energy and Pressure}

We calculate the energy of the system in the leading order of the hole-line expansion, 
that is we replace Green's functions in Eq.(\ref{eq:ene_exp}) with 
the free one $G^{f,b}_0$ in Eq.(\ref{eq:freeGF}), and obtain 

\begin{align}
        E \sim - iV \int \frac{d^4 q}{(2\pi)^4} \left( \epsilon^f_{\bf q} 
                     + \Sigma^f(q) \right) G^f_0(q) \textrm{e}^{iq_0\eta} \nonumber \\
         = E_0N_f + N_0 \int \frac{d^3 {\bf p}}{( 2\pi )^3} \theta ( k_F-| {\bf p}| )
                \Gamma(\epsilon^f_{\bf p}, {\bf p}) ,
                \label{eq:e_no_bb}
        \end{align}
where, $E_0 = 3/5E_F$ and $E_F$ is the Fermi energy. Within the same 
approximations, the thermodynamic potential at zero temperature is given by
\begin{equation}
        \Omega = E_0N_f + N_0 \int \frac{d^3 {\bf p}}{\left( 2\pi \right)^3}
         \theta ( k_F -| {\bf p} | )\Gamma(\epsilon^f_{\bf p}, {\bf p}) - \mu_b N_0 .
                \label{eq:Omegaeq}
        \end{equation}
Thus, the equilibrium condition (\ref{eq:bnum}) for $\Omega$  leads 
to the integral equation for $\mu_b$,
\begin{align}
        \mu_b  = \int \frac{d^3{\bf p}}{(2\pi)^3} \theta ( k_F - | {\bf p}| )
                \Gamma(\epsilon^f_{\bf p}, {\bf p}),
                 \label{eq:chemdet}
      \end{align}
where $\Gamma$ depends also on $\mu_b$.
The total energy of the system is then finally 
given by 
\begin{equation}
        E  = E_0N_f + \mu_b N_b.
        \label{eq:intene} 
    \end{equation}
Details of the calculation and the analytic expression for $\Gamma$ are given in the appendix. 
 

We may rewrite Eq.(\ref{eq:chemdet}) in a scaled form as
\begin{equation}
    \tilde{\mu}_b = 2 \left( 1+\frac{1}{\zeta} \right)  \int^1_0 d\tilde{p} \, 
    \tilde{p}^2 \tilde{\Gamma}(\tilde{p},\tilde{\mu}_b,\zeta) ,
    \label{eq:iona}
    \end{equation}
where we introduced tilde (dimensionless) quantities 
through $\Gamma(p, \mu_b) = 2\pi^2/\nu k_F \tilde{\Gamma}(\tilde{p}, \tilde{\mu}_b)$, 
$\tilde{a} = k_F a$, $\tilde{p} = |{\bf p}|/k_F$, $\tilde{\mu}_b = \mu_b / E_F$. 
The expression shows that the scaled chemical potential $\tilde{\mu}_b$ depends only 
on the mass ratio $\zeta=m_b/m_f$ and the dimensionless scattering length $\tilde{a} = k_F a$.
We solved Eq.(\ref{eq:iona}) for $\tilde{\mu}_b$  numerically as a function of the 
boson-fermion mass ratio  $\zeta$ for different values of the interaction 
strength represented by $\tilde{a}$. In terms of the scaled quantities, the ground state
energy per particle is expressed from Eq.(\ref{eq:intene})  as
\begin{equation}
        \frac{E}{N_f}=\frac35 E_F ( 1 + \beta ),
         \label{eq:aa}
     \end{equation}
where the  dimensionless  parameter $\beta$ is given by 
\begin{equation}
        \beta = \frac53 \tilde{\mu}_b \frac{N_b}{N_f} .
       \end{equation}
       
We first show the results for energy and pressure in the unitarity limit, 
$|a| \rightarrow \infty$. In this limit, assuming S-wave scattering  and neglecting 
effective range, we are left with only one length scale, $k_F^{-1}$, or $n_f^{-1/3}$ in 
terms of the density $n_f=N_f/V$ \cite{unit_th,univ_hyp} for a given mass ratio $\zeta$. 
Note that the chemical potential $\tilde{\mu}_b$ has no $k_F$ 
dependence in the unitarity limit, and the parameter $\beta$ depends only 
on the number ratio $N_b/N_f$. Thus the ground state energy per particle,  
Eq.(\ref{eq:aa}), is  proportional to $E_F$, and the dependence on the 
parameters are all absorbed in a simple multiplicative factor $(1+\beta)$.

%
\begin{figure}[htb]
  \centering
  \rotatebox{-90}{%
   \includegraphics[scale=0.45]{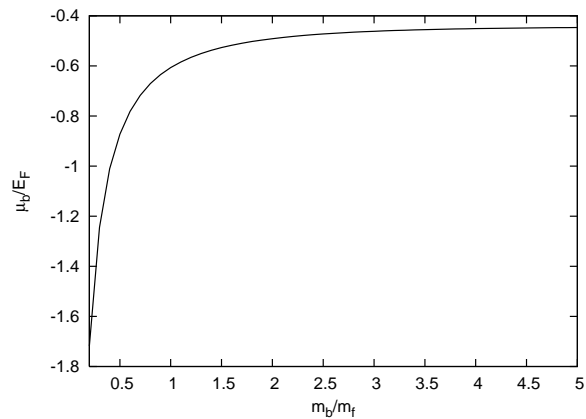}
   }
   \caption{Scaled chemical potential $\tilde{\mu}_b=\mu_b/E_F$ as a function 
   of the boson-fermion mass ratio $\zeta=m_b/m_f$ in the unitarity limit.}
   \label{fig:chem}
\end{figure}
\begin{figure}[htb]
   \centering
   \rotatebox{-90}{%
   \includegraphics[scale=0.45]{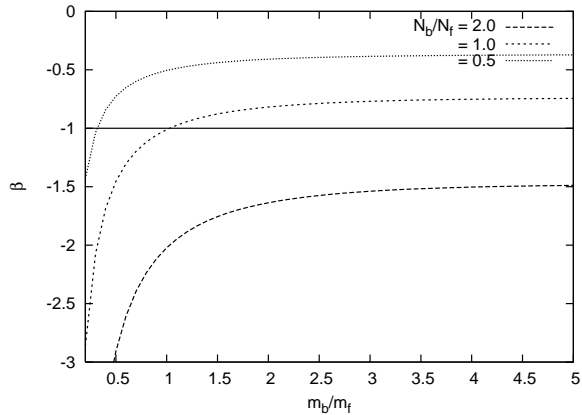}
  }
   \caption{$\beta$ parameter as a function of the boson-fermion mass ratio 
   in the unitarity limit 
   for $N_b/N_f=2.0,1.0,0.5$ from top to bottom.}
   \label{fig:graene}
\end{figure}
%

We show in  fig.\ref{fig:chem} the chemical potential $\mu_b$ and in 
fig.\ref{fig:graene} the beta parameter as functions of the mass ratio $\zeta=m_b/m_f$,
both in the unitarity limit. 
Note that the results are independent of the magnitude of the individual mass 
parameters as we show dimensionless quantities scaled with $E_F$. 
Figure \ref{fig:chem} shows that the boson chemical 
potential is always negative. This fact reflects the attractive boson-fermion 
interaction in the unitarity limit, in accordance with  Eq.(\ref{eq:renorm}) which 
implies negative $g_{bf}$. The behavior of $\beta$ in 
fig.\ref{fig:graene} simply follows the one of the chemical potential. The results 
suggest that the attractive interaction becomes more effective for small values of 
the mass ratio $\zeta$, and the effect is greatly enhanced as the particle number 
ratio $N_b/N_f$ becomes larger, 
that is as the number of bosons increases with respect to the fermions.
The dependence on $\zeta$ may partly  be understood 
by noting that the relative phase space available for the intermediate states 
in the two-body scattering in the mixture will be larger for small $\zeta$, i.e., 
for a relatively larger $m_f$, because of the lower Fermi energy and higher level 
density. 

We next consider the pressure  to study the stability of the system. 
Since the total energy takes a universal form and
the $\beta$ parameter has no volume dependence in the unitarity limit, 
the pressure is simply given by
\begin{equation}
        P = \frac{\partial E}{\partial V} = \frac25 \frac{N_f}{V} E_F ( 1+\beta ) 
        \qquad \hbox{(unitarity limit)}
        \label{eq:pre}.
        \end{equation}
For $\beta< -1$ the pressure becomes negative, and the system 
collapses. This happens especially for larger values of $N_b/N_f$, where the 
pressure becomes always negative irrespective of the mass ratio $\zeta$. 
In actual experiments, e.g., for the $^{40}$K-$^{87}$Rb mixture which has 
$\zeta \sim 2.3$, the number ratio is typically $\mathcal{O}(1)\sim\mathcal{O}(10^3)$ 
and the system would collapse in the unitarity limit. This is not in contradiction 
to recent experimental results \cite{BFcollapse}.

Another feature in the unitarity limit seen from fig.\ref{fig:chem} and 
fig.\ref{fig:graene} is that the boson chemical potential and the ground state energy 
saturate when the mass ratio becomes large. It is natural that the bosonic degree 
of freedom gets frozen and a universal fermionic description appears in this case, 
since the fermionic effects on the bosons would become negligible, and the bosons 
would act as a static external field for fermions. This behavior at large $\zeta$ 
also has been discussed in \cite{albus}. 

\begin{figure}[htb]
  \centering
  \rotatebox{-90}{%
   \includegraphics[scale=0.45]{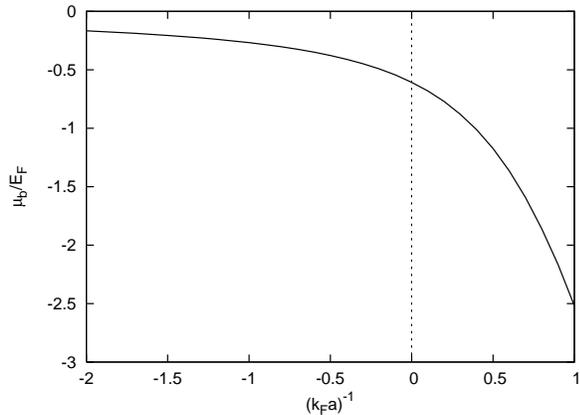}
   }
   \caption{Scaled chemical potential $\tilde{\mu}_b=\mu_b/E_F$ as a function 
   of $(k_Fa)^{-1}$ for $m_b/m_f=N_b/N_f=1$.}
   \label{fig:chem_a}
\end{figure}
\begin{figure*}[htb]
 \begin{minipage}[b]{0.4\hsize}
  \begin{center}
   \rotatebox{-90}{%
   \includegraphics[scale=0.4]{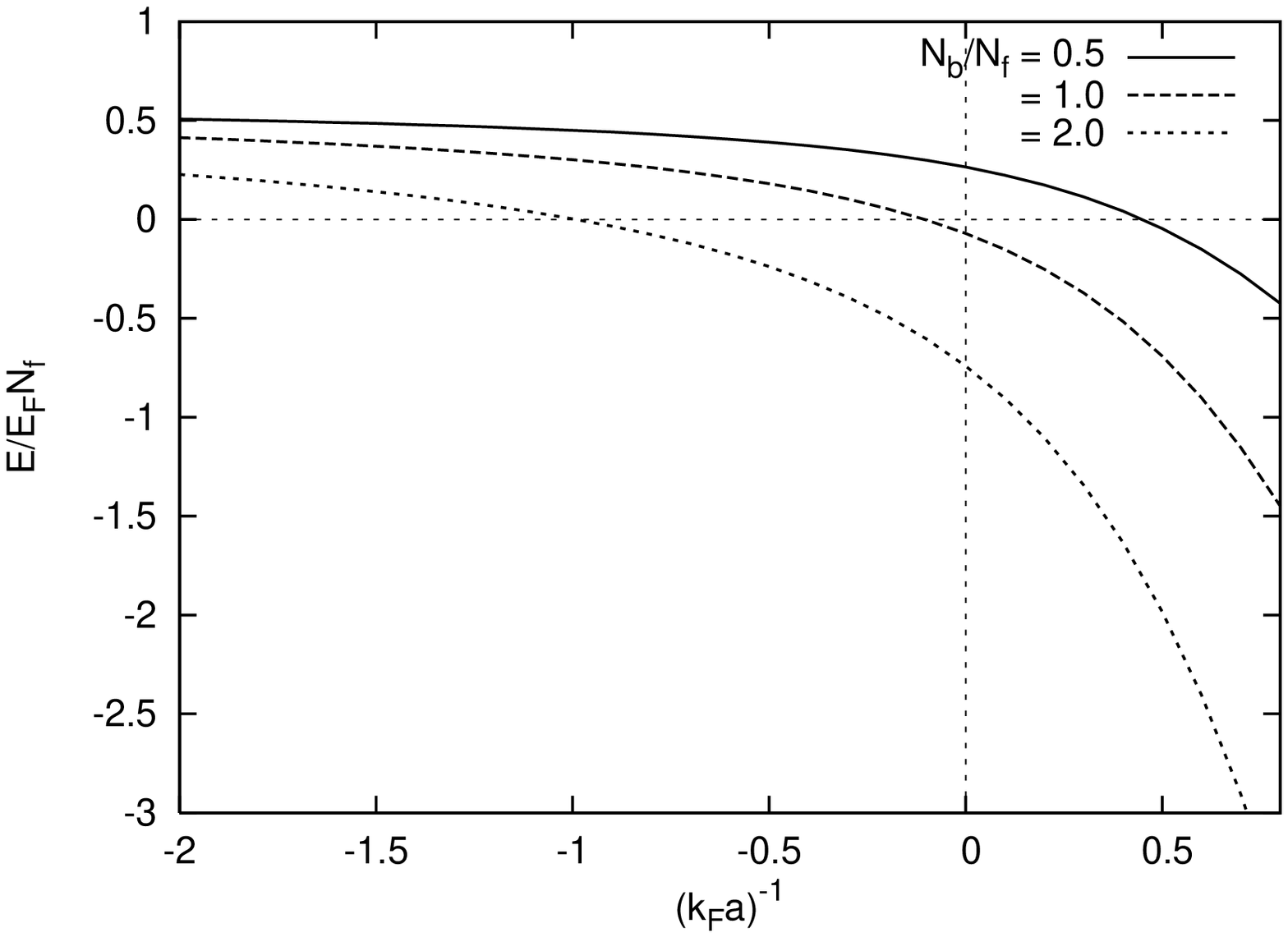}
   }
  \end{center}
 \end{minipage}
 \hspace{5mm}
 \begin{minipage}[b]{0.4\hsize}
  \begin{center}
   \rotatebox{-90}{%
   \includegraphics[scale=0.4]{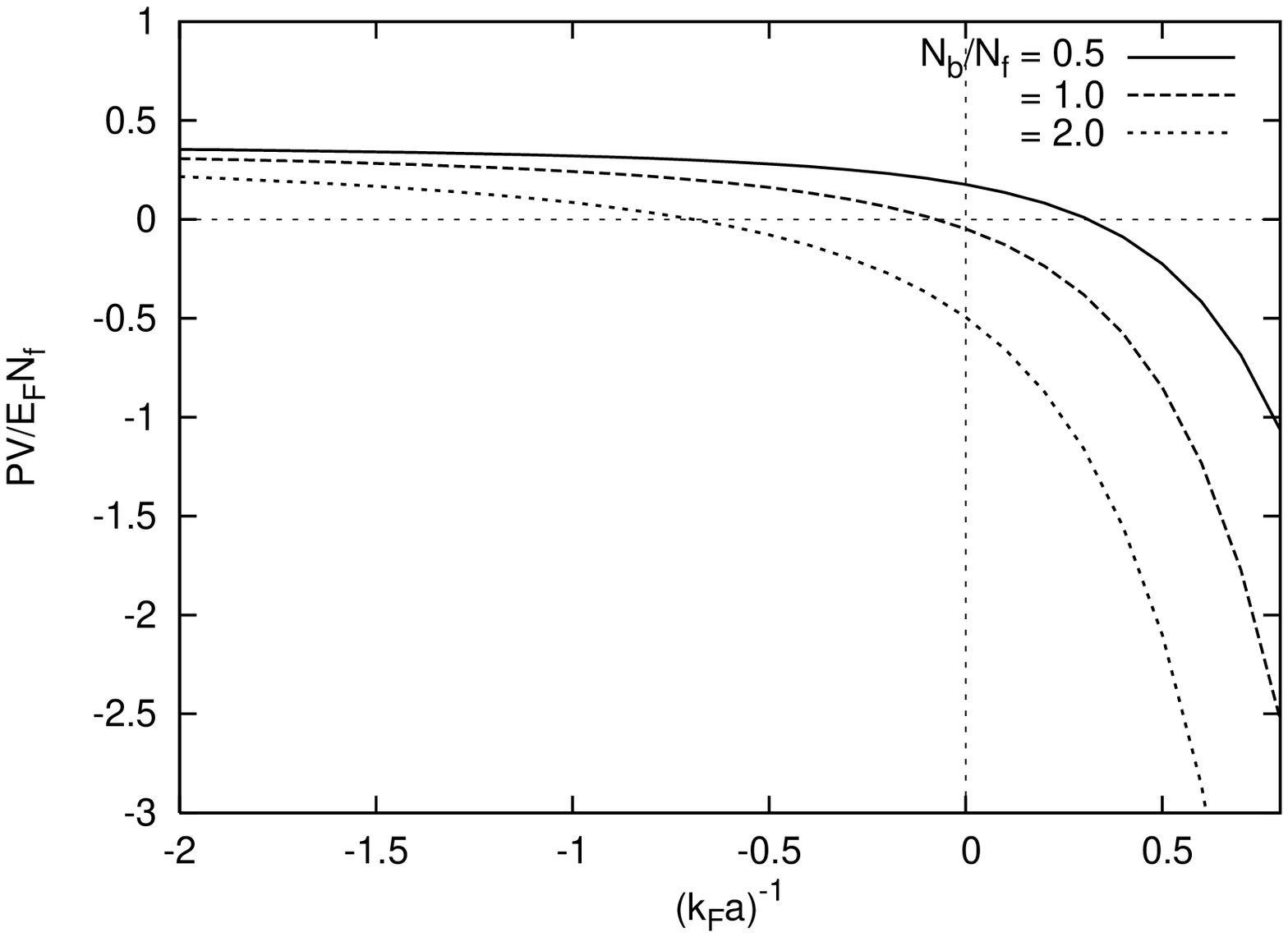}
   }
  \end{center}
 \end{minipage}
        \caption{Scaled energy $E/E_FN_f$ (left panel) and pressure $PV/E_FN_f$ 
        (right panel) as a function of  $(k_Fa)^{-1}$ for $m_b/m_f=0.8$.
         Three curves corresponds to $N_b/N_f=0.5, 1.0, 2.0$ from top to bottom.
         The vertical line indicates the unitarity limit $a=\infty$.}
    \label{fig:energy_pressure_m08}
\end{figure*}
\begin{figure*}[htb]
 \begin{minipage}[b]{0.4\hsize}
  \begin{center}
   \rotatebox{-90}{%
   \includegraphics[scale=0.4]{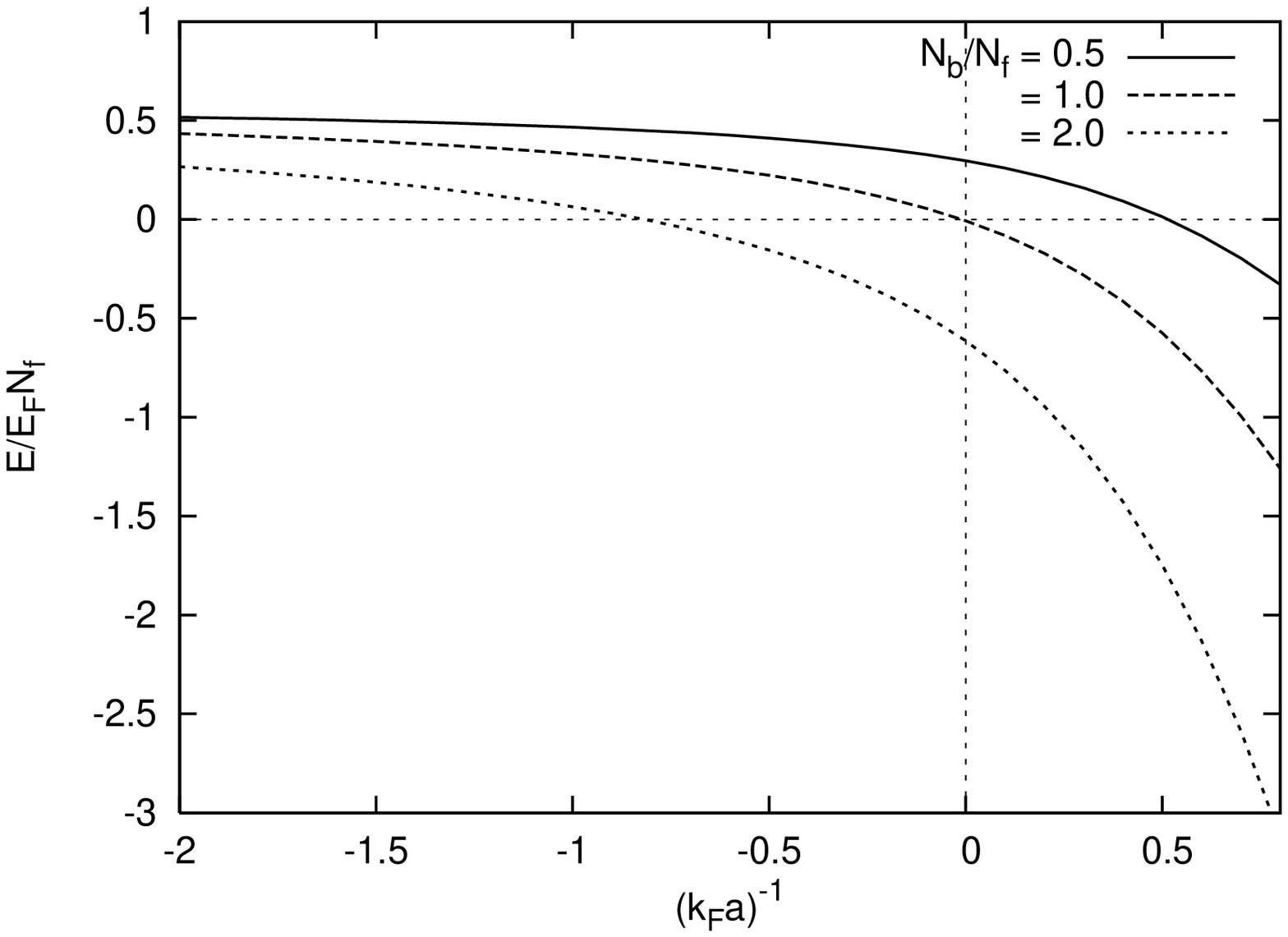}
   }
  \end{center}
 \end{minipage}
 \hspace{5mm}
 \begin{minipage}[b]{0.4\hsize}
  \begin{center}
   \rotatebox{-90}{%
   \includegraphics[scale=0.4]{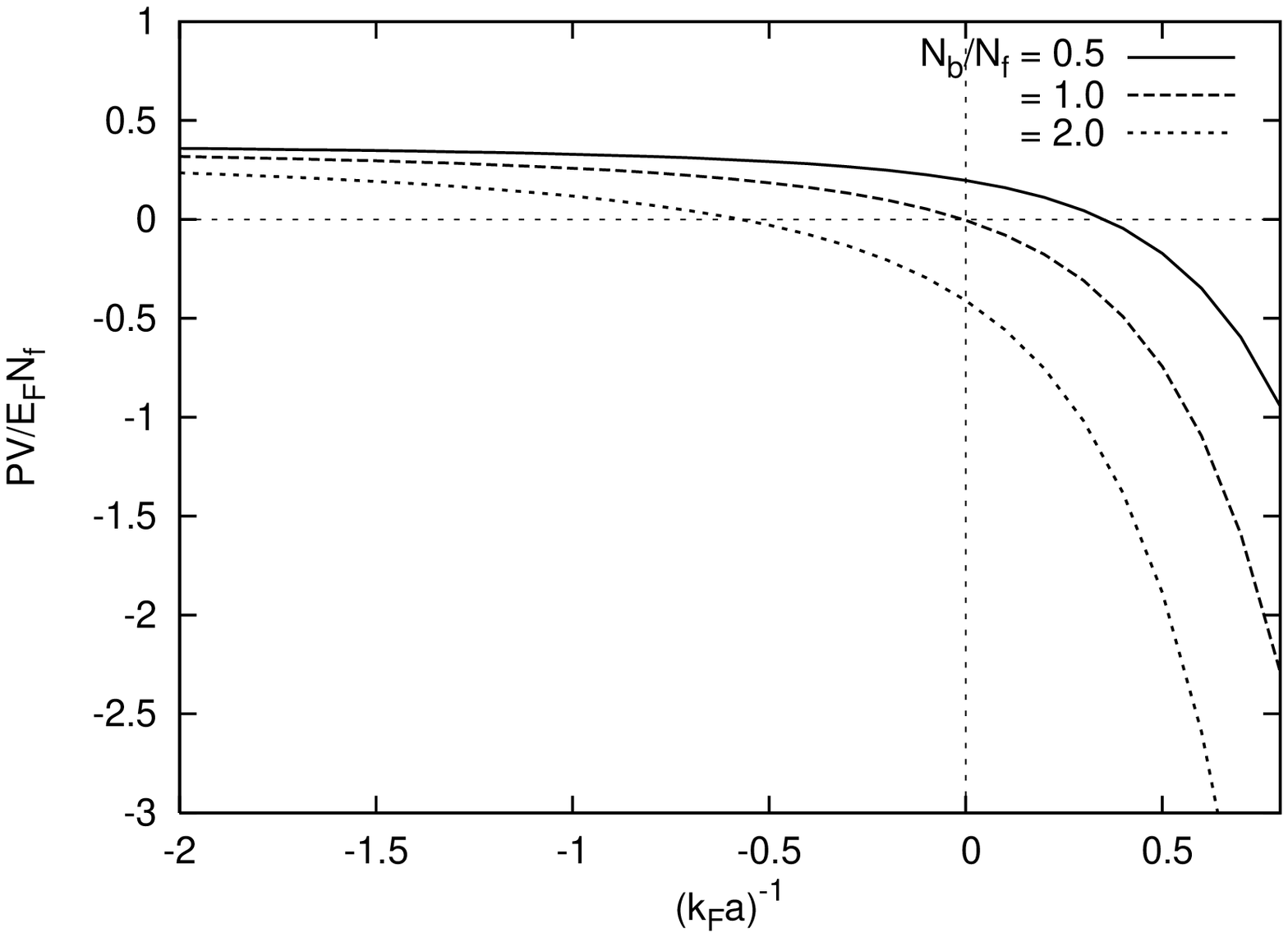}
   }
  \end{center}
 \end{minipage}
        \caption{Same as fig.\ref{fig:energy_pressure_m08}, but for $m_b/m_f=1.0$.}
    \label{fig:energy_pressure_m10}
\end{figure*}
\begin{figure*}[htb]
 \begin{minipage}[b]{0.4\hsize}
  \begin{center}
  \rotatebox{-90}{%
   \includegraphics[scale=0.4]{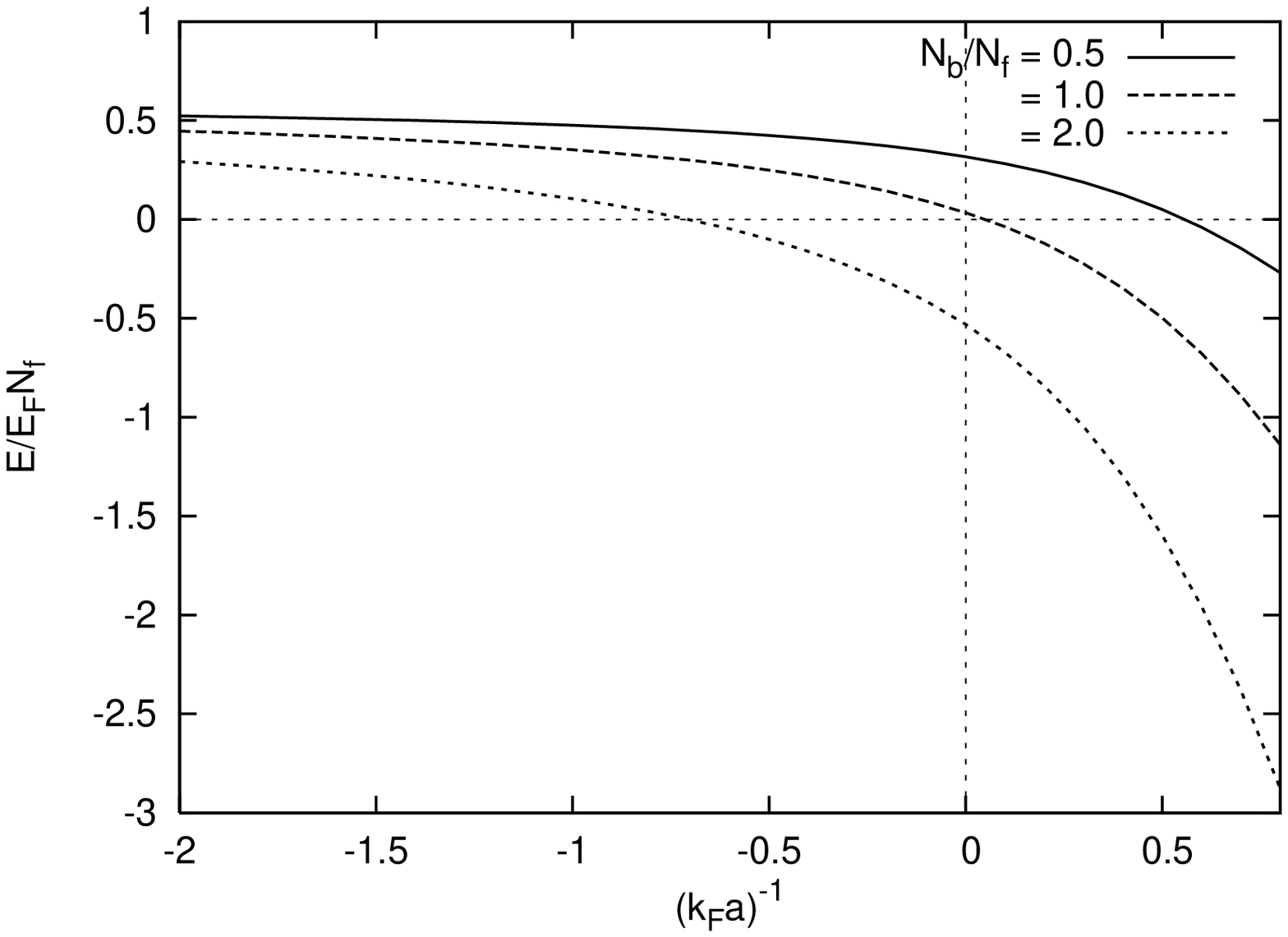}
   }
  \end{center}
 \end{minipage}
 \hspace{5mm}
 \begin{minipage}[b]{0.4\hsize}
  \begin{center}
   \rotatebox{-90}{%
   \includegraphics[scale=0.4]{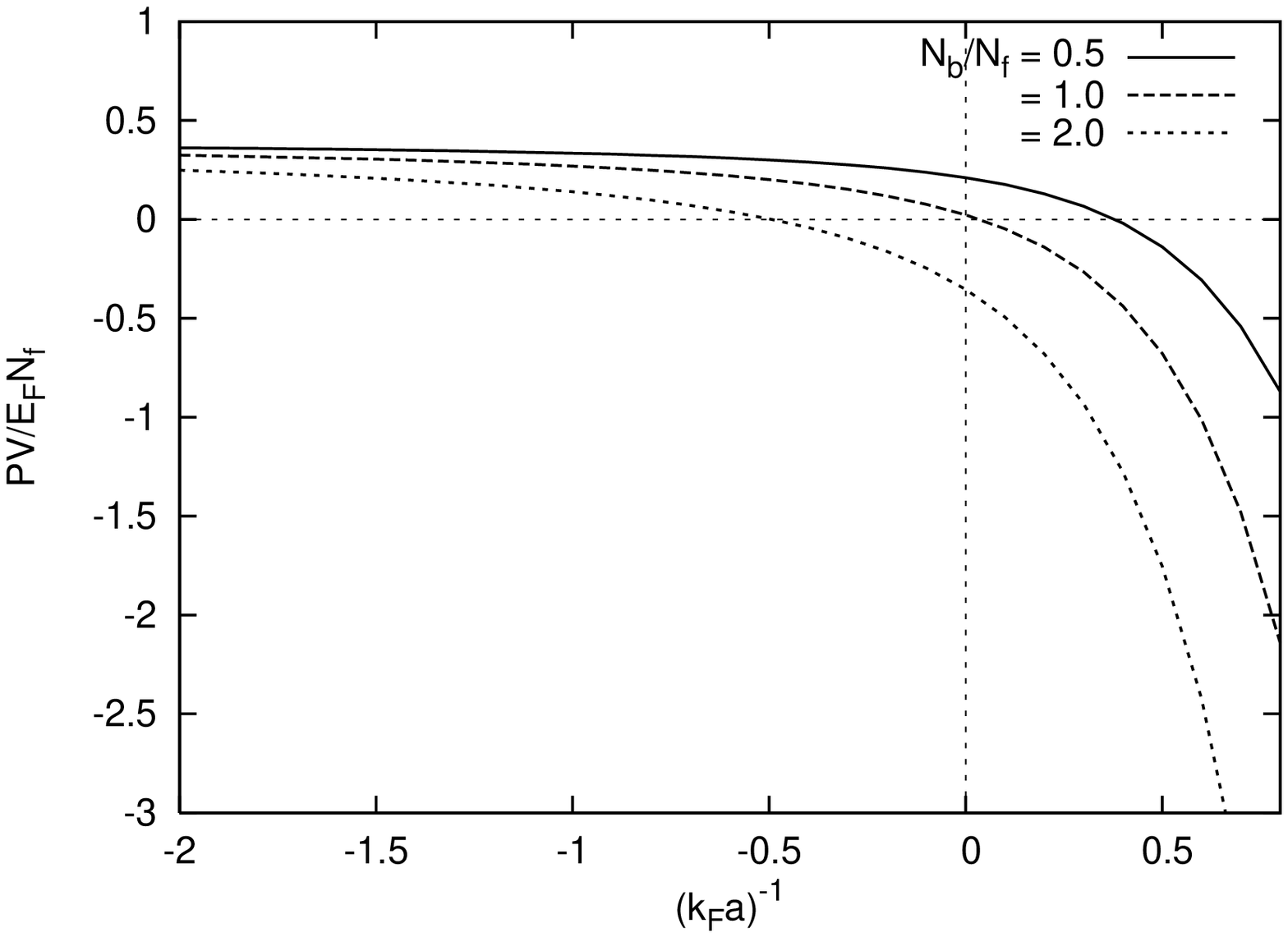}
   }
  \end{center}
 \end{minipage}
        \caption{Same as fig.\ref{fig:energy_pressure_m08}, but for $m_b/m_f=1.2$.}
    \label{fig:energy_pressure_m12}
\end{figure*}

We now study the case with an arbitrary value of the scattering length. 
First, we show the chemical potential $\mu_b$ as a function of $(k_Fa)^{-1}$ in fig.\ref{fig:chem_a}, 
where the mass ratio $\zeta$ and $N_b/N_f$ are set to 1.
Then, we show the 
energy and pressure as a function of $(k_Fa)^{-1}$ in figs.\ref{fig:energy_pressure_m08},
\ref{fig:energy_pressure_m10} and \ref{fig:energy_pressure_m12}, for the mass ratio 
$\zeta=0.8,1.0,1.2$ with different values of $N_b/N_f=0.5, 1.0, 2.0$. For $a \neq 0$, 
the pressure is given by 
\begin{widetext}
\begin{equation}
        P = \frac25 \frac{N_f}{V} E_F \left[ 1+ \frac53 \left\{
             2 \left( 1+\frac{1}{\zeta} \right)  \frac{N_b}{N_f}  \int^1_0 d\tilde{p} 
             \, \tilde{p}^2  \left( \tilde{\Gamma}(\tilde{p},\tilde{\mu}_b,\zeta)
             - \frac{\pi}{\tilde{a}} \tilde{\Gamma}^2(\tilde{p},\tilde{\mu}_b,\zeta)
                  \right) \right\} \right] ,
        \label{eq:press_general_a}
\end{equation}
\end{widetext}
where the term dependent on $\Gamma^2$ reflects that the chemical potential, and 
hence the $\beta$ parameter, depends on $V$. From the results on energies we see that the 
system becomes more attractive at smaller values of the mass ratio $\zeta$ as in the 
unitarity limit, although the effect is not large in this parameter range. 
We find a strong increase of the attraction  as  the parameter $(k_Fa)^{-1}$ passes 
through zero, the unitarity limit, from negative to positive. This is in accord with 
a naive picture where the positive values of the scattering length imply a newly 
formed bound state. One should however note that even in the present approximation 
the effects of the medium modify the two-body scattering amplitude, and a simple picture 
of independent bound pairs does not hold in general. Turning now to the pressure,  a 
comparison of Eq.(\ref{eq:press_general_a}) with the corresponding expression 
(\ref{eq:pre})  in the unitarity limit shows that the strong attraction for $a>0$ 
comes from the large negative values of $\Gamma$ as well as from the coherence of 
the two terms in the integrand. 


\section{Bose-Fermi Pair Correlation}

Results of the previous section indicate that the strong attraction in the mixture will 
show up especially for positive $a$, which may eventually lead to a collapse of 
the system. We now consider another scenario 
for the attractively interacting mixture, the possibility of a Bose-Fermi pair 
formation\cite{schuck,bf_pair}. For this purpose we study in this section the 
behavior of the pole of the Bose-Fermi scattering amplitude $\Gamma(P)$ in the 
mixture. 

In unpolarized (or two-component)  Fermi systems, an infinitesimal attraction around 
the Fermi surface leads to formation of Cooper pairs with a center-of-mass (CM) 
momentum $\bf P=0$, 
causing a transition to the BCS state. In a Bose-Fermi mixture, on the other hand, 
the difference in the momentum distribution of the two particles and the fermionic 
character of the Bose-Fermi pair, in particular, predict quite a different scenario 
for the formation of the pairs in the mixture. It requires  consideration of the 
balance of the kinetic energies of different kinds of particles and the magnitude 
of the attractive interaction.

\subsection{Preliminary considerations}

Let us  give a brief picture on the formation of Bose-Fermi pairs in the 
mixture. Following the idea of \cite{schuck}, we may take a 
Bose-Fermi mixture with $N_b=N_f$, focusing only one pair of a boson and a fermion with CM momentum 
$\bf P$, putting other particles as a free background. 
The Hamiltonian in this system may 
be written as
\begin{equation}
    H=\frac{{\bf p}_b^2}{2m_b}+\frac{{\bf p}_f^2}{2m_f}+V_{bf}+H_{\rm bg}, \quad
    ({\bf p}_b+{\bf p}_f={\bf P})
    \label{eq:model_H}
    \end{equation}
where we explicitly write the kinetic energy and the interaction for the two particles, 
while the background Hamiltonian $H_{\rm bg}$ acts only to impose a Pauli-principle constraint. 
We neglect here the effect of the boson chemical potential for simplicity. 
If  the effect of $V_{bf}$ on the pair were negligible, the energy of the Bose-Fermi pair with 
CM momentum $\bf P$ would be simply 
\begin{equation}
        \epsilon_{\rm free}(|{\bf P}|)=\frac{{\bf P}^2}{2m_f},
        \label{eq:eps_free}
        \end{equation}
which has been called {\it free branch} in \cite{schuck}, since the boson will  remain at 
${\bf p}_b=0$ in the condensate.  When the effect of the interaction becomes important, one 
may rewrite  Eq.(\ref{eq:model_H}) as
\begin{equation}
    H=\frac{\bf P^2}{2(m_b+m_f)}+H_{\rm rel}+H_{\rm bg},
    \label{eq:model_Hrel}
    \end{equation}
where the interaction $V_{bf}$ is contained in the Hamiltonian $H_{\rm rel}$ of the relative 
motion. By replacing the latter  with its 
eigenvalue $E_{\rm rel}$ (the effect of the medium may be included here), one 
obtains a different dispersion curve for the Bose-Fermi pair, the {\it collective branch},
which is given by
\begin{equation}
       \epsilon_{\rm coll}(|{\bf P}|)=\frac{\bf P^2}{2(m_b+m_f)}+E_{\rm rel}.
       \label{eq:eps_coll}
       \end{equation}
The calculation for the model with a separable interaction in \cite{schuck} shows 
that the two dispersion curves, Eqs.(\ref{eq:eps_free}) and (\ref{eq:eps_coll}),
coexist except for the region of CM momentum ${\bf P}_c$ given by a solution 
of the equation 
\begin{equation}
     \epsilon_{\rm free}(|{\bf P}_c|)=\epsilon_{\rm coll}(|{\bf P}_c|).
     \label{eq:Pcross}
     \end{equation}
There is a mixture of the two branches around $|{\bf P}|=P_c$, and the dispersion curve of the 
pair deviates from $\epsilon_{\rm free}$ and $\epsilon_{\rm coll}$. 
For a sufficiently attractive interaction, the solution of Eq.(\ref{eq:Pcross}) satisfies the 
condition $|{\bf P}_c|\leq k_F$, which occurs for
\begin{equation}
     E_{\rm rel}\leq \frac{m_b}{m_b+m_f}E_F.
     \label{ed:cond_cross}
     \end{equation}
One may thus expect that the system will lower the energy by converting the free Bose-Fermi 
particles in the range $|{\bf P}_c|\leq |{\bf P}|\leq k_F$ into {\it collective} pairs. 
(Note that we neglect 
the interaction of the pairs in this simple argument.) If $V_{bf}$ is so strong as to 
allow for a bound state of the 
pair, i.e., $E_{\rm rel}<0$, all the free bosons and fermions would be 
replaced with the bound B-F pairs, and the new Fermi sea of the pairs will be formed. 

This picture may be modified even in this simple model, however, for a system with many 
Bose-Fermi pairs. As the pairs can occupy low-momentum states having lower energies 
without changing  the total momentum of the system,  the formation of the collective 
pairs may start if the condition 
\begin{equation}
     E_{\rm rel}=\epsilon_{\rm coll}(|{\bf P}|=0)\leq \epsilon_{\rm free}(k_F)=E_F
     \end{equation}
is satisfied, even before the condition (\ref{ed:cond_cross}). Similarly, the 
formation of the Fermi sea of the pair will be completed only when 
\begin{equation}
      \epsilon_{\rm coll}(k_F) \leq\epsilon_{\rm free}(|{\bf P}|=0)=0 
      \end{equation}
is satisfied. 

We note that the collective branch (\ref{eq:eps_coll}) may suffer a Laudau type 
damping for sufficiently large $E_{\rm rel}$. This is because the free 
Bose-Fermi mixture with  Fermi momentum $k_F$ has a continuum of the boson-fermion 
excitation with momentum ${\bf P}={\bf p}_b+{\bf p}_f$ which starts at the energy \cite{schuck}
\begin{equation}
     \epsilon_{\rm th}(|{\bf P}|)=\frac{(|{\bf P}|-k_F)^2}{2m_b}+E_F.
     \label{eq:eps_th}
     \end{equation}
This implies that the collective branch remains undamped only when 
\begin{equation}
        \epsilon_{\rm coll}(|{\bf P}|)\leq \epsilon_{\rm th}(|{\bf P}|),
        \end{equation}
a condition which is approximately realized for the pole of the boson-fermion 
scattering amplitude as shown below.

\subsection{Pole behavior of the two-particle scattering amplitude}

\begin{figure}[htb]
        \begin{center}
	\rotatebox{-90}{
        \includegraphics[scale=.45]{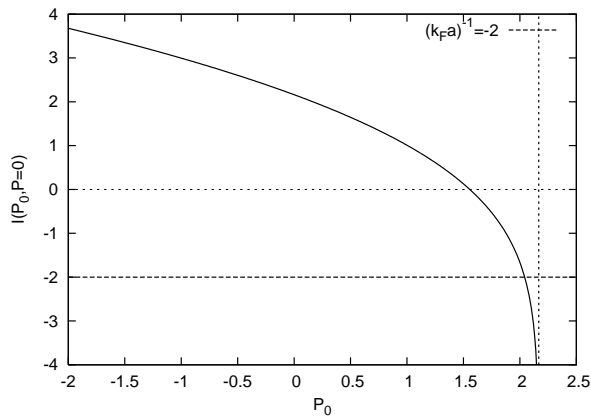}
	}
        \end{center}
        \caption{The behavior of the right hand side of eq.(\ref{eq:pole_condition}). 
			The horizontal line is the left hand side. Here, $m_b/m_f=1$.
			The vertical line shows the the position of the logarithmic divergence.}
        \label{fig:I_by_p_zero}
\end{figure}

Now we study the behavior of the pole of $\Gamma(P)$. 
A first study concerns the pole condition.
That is 
\begin{equation}
        \frac{\nu}{2\pi a} = I\left(P_0, \left|{\bf P}\right|\right) \label{eq:pole_condition}.
\end{equation}
For $|{\bf P}|=0$, we show in fig.\ref{fig:I_by_p_zero} the right hand side of (\ref{eq:pole_condition}) 
as a function of $P_0$ for $m_b/m_f=1$. 
We see the development of a logarithmic divergency as $P_0$ aproaches $k_F^2/2\nu - \mu_B$. 
This stems from the fact that above dispersion integral has exactly the same structure as
the one encountered in the problem of Cooper for a fermion pair in a Fermi-sea. 
We want, however, to point out that the pole corresponds to a composite fermion 
what has important consequences for the physics. 
Nevertheless the fact is there that a stable collective B-F pair developes for any 
infinitesimal attraction, i.e, even in the limit $a\rightarrow -0$ quite in analogy to the 
original Cooper pole \cite{fetter}.

\begin{figure}[htb]
        \begin{center}
        \includegraphics[scale=.5]{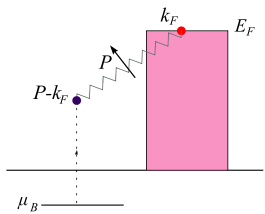}
        \end{center}
        \caption{Rough sketch of the formation of the Bose-Fermi pair with CM momentum $\bf P$.
                        The dots represent a boson (left) and a fermion(right). 
                        The shaded box indicates a fermi sphere.}
        \label{fig:pairing}
\end{figure}

Let us now discuss the collective pole contained in $\Gamma(P)$.
Be $P_0^c$ the pole of $\Gamma(P)$ with CM momentum ${\bf P}$. 
$P_0^c(|{\bf P}|)$ represents the total energy of a boson-fermion pair corresponding to the 
{\it collective branch} aside from the chemical potential. 
One may then define a $|{\bf P}|$-dependent binding energy (including total kinetic energy) 
measured from the last filled free Bose-Fermi pair by
\begin{equation}
        \Delta_{pair}(|{\bf P}|) =  \epsilon_{\rm free}(k_F)  - \mu_b - P_0^c(|{\bf P}|),
        \label{eq:be} 
\end{equation}
see fig.\ref{fig:pairing}.
Positive value of $\Delta_{pair}(|{\bf P}|)$ would signal a formation of the Bose-Fermi pair. 

\begin{figure*}[htb]
 \begin{minipage}[b]{0.4\hsize}
  \begin{center}
   \rotatebox{-90}{%
   \includegraphics[scale=0.4]{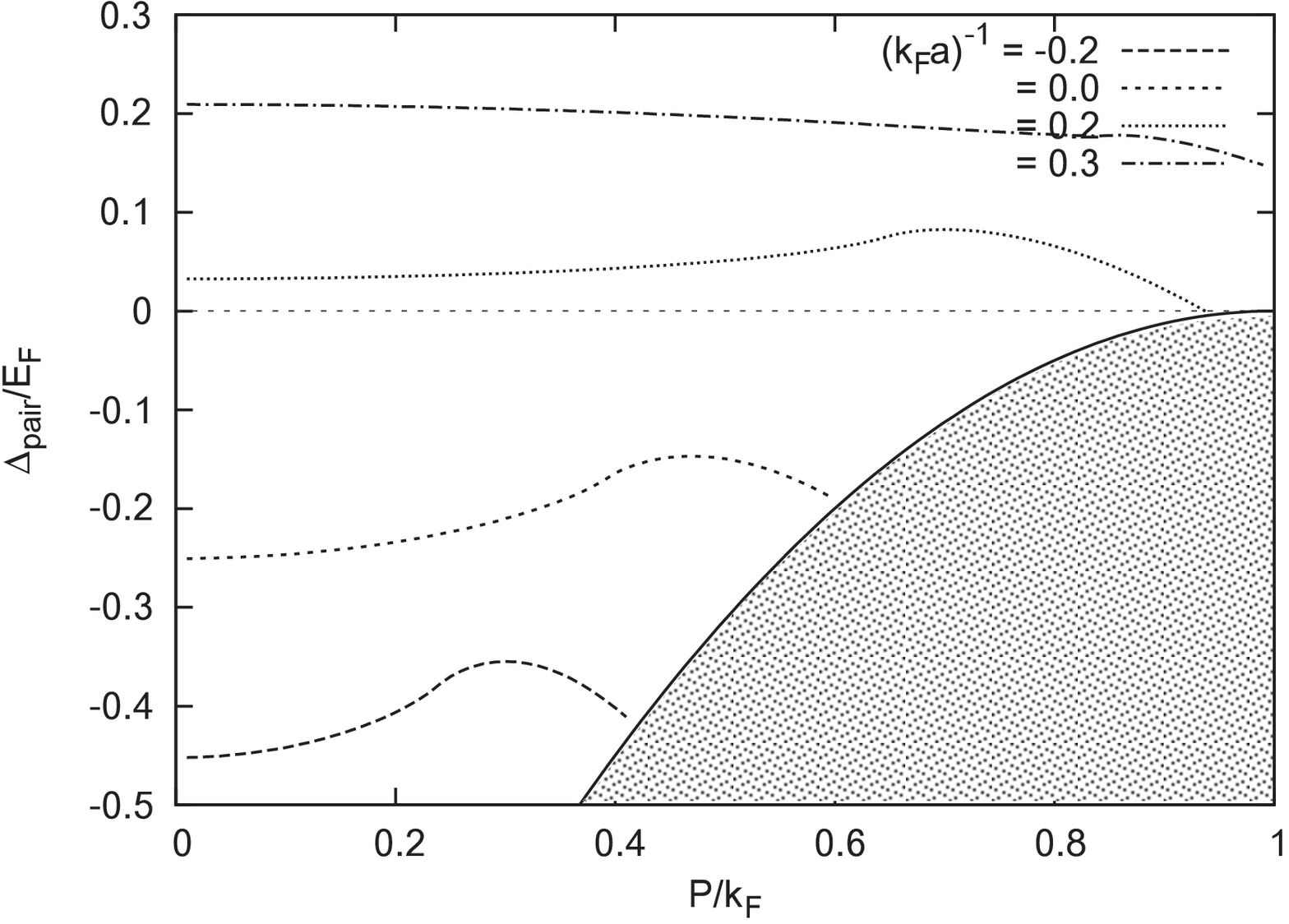}
   }
  \end{center}
 \end{minipage}
 \hspace{5mm}
 \begin{minipage}[b]{0.4\hsize}
  \begin{center}
  \rotatebox{-90}{%
   \includegraphics[scale=0.4]{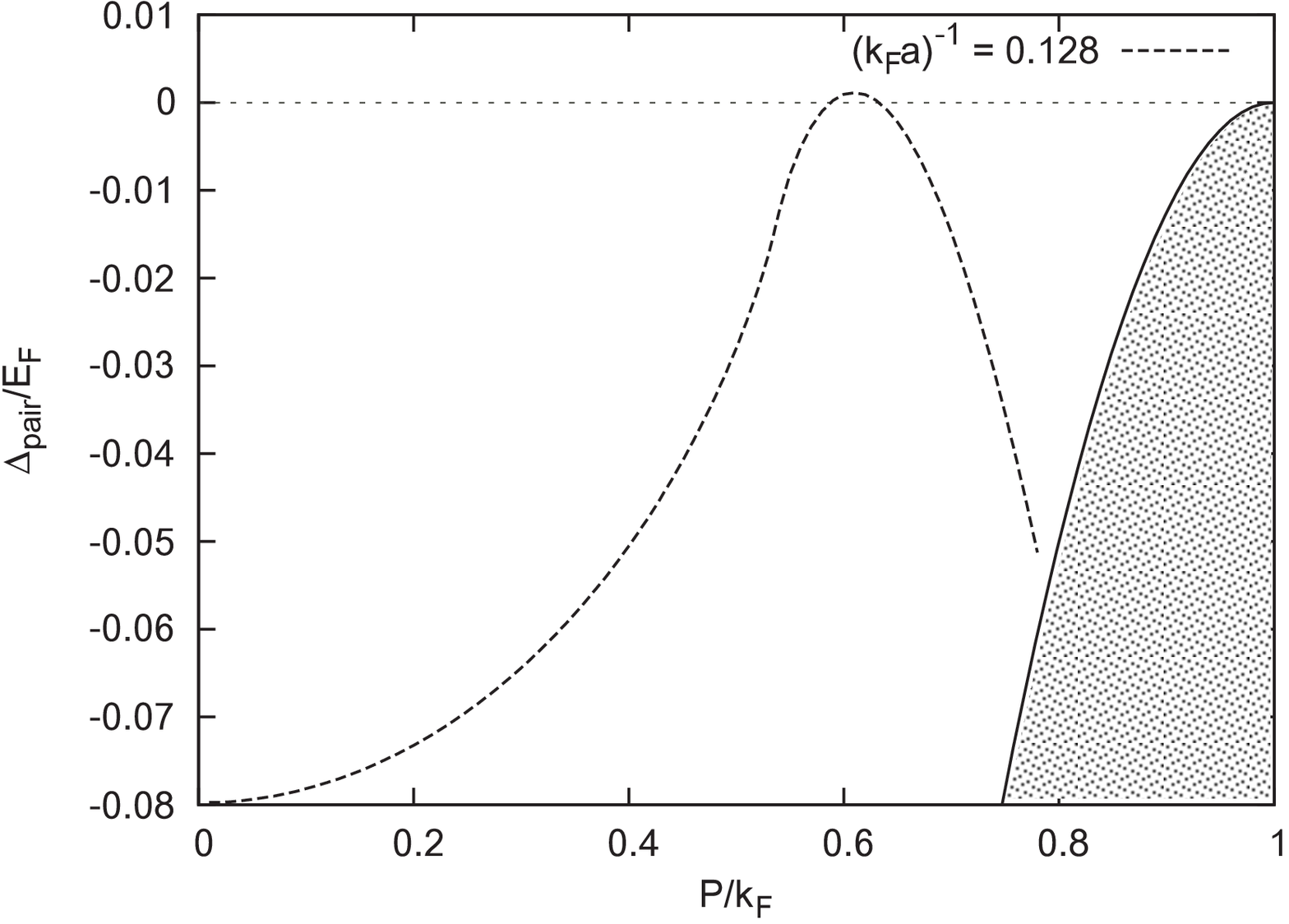}
   }
  \end{center}
 \end{minipage}
  \caption{$\Delta_{pair}(P)$ as a function of center of mass momentum $|{\bf P}|$ for
                various $(k_Fa)^{-1}$ with $m_b/m_f = 0.8$ and $N_b/N_f=1$. The shaded area shows the 
                contiuum of the excitation which starts at $\epsilon_{\rm th}$, 
                Eq.(\ref{eq:eps_th}). The right  panel shows the 
                enlarged figure at the threshold value of $(k_Fa){-1}$ (see text). 
                Note the scale difference 
                in the vertical axis between left and right panels.}
 \label{fig:delta_pair_m08}
\end{figure*}

\begin{figure*}[htb]
 \begin{minipage}[b]{0.4\hsize}
  \begin{center}
   \rotatebox{-90}{%
   \includegraphics[scale=0.4]{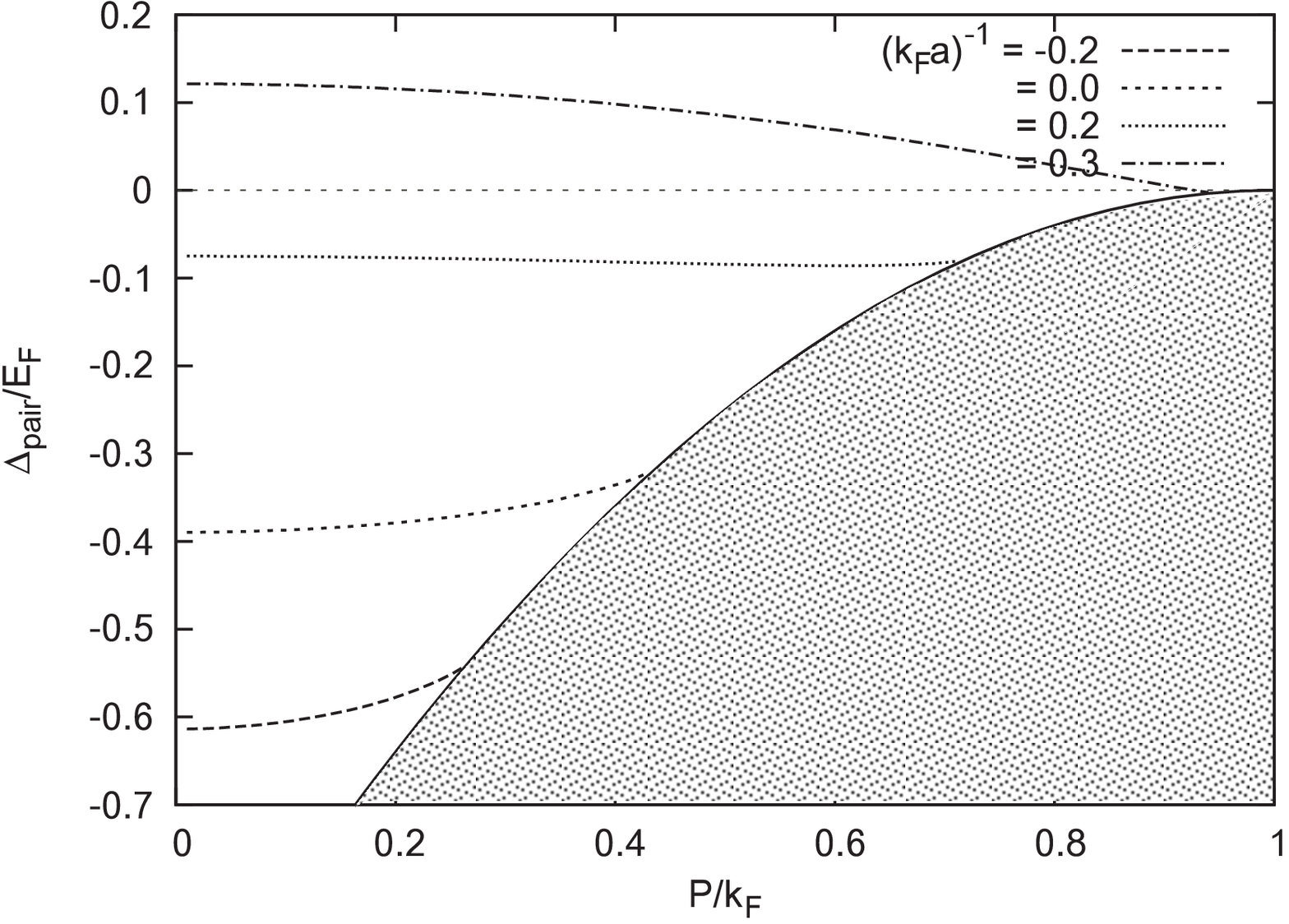}
   }
  \end{center}
 \end{minipage}
 \hspace{5mm}
 \begin{minipage}[b]{0.4\hsize}
  \begin{center}
   \rotatebox{-90}{%
   \includegraphics[scale=0.4]{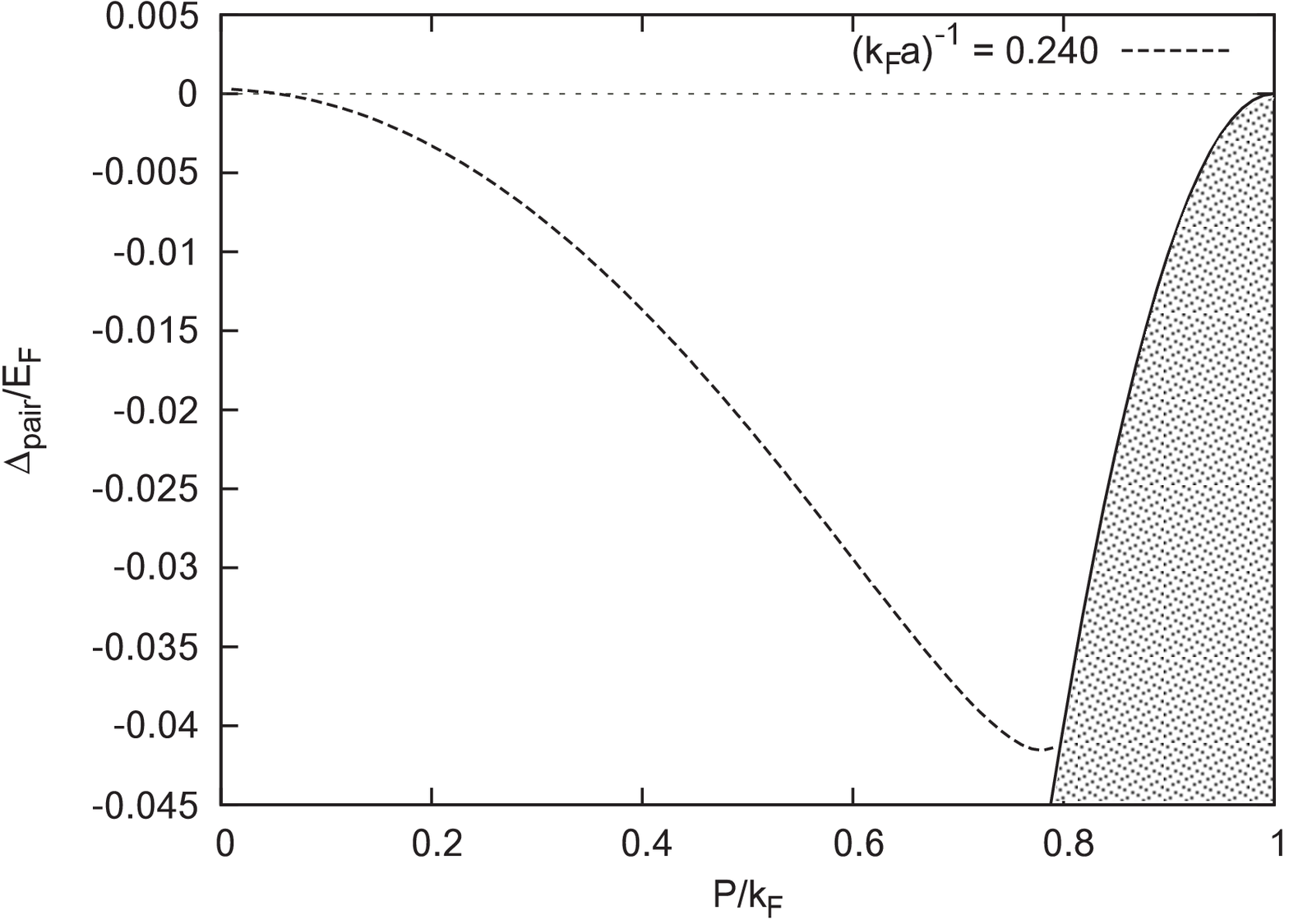}
   }
  \end{center}
 \end{minipage}
  \caption{Same as Fig.\ref{fig:delta_pair_m08} with $m_b/m_f = 1.0$.}
  \label{fig:delta_pair_m10}
\end{figure*}

\begin{figure*}[htb]
 \begin{minipage}[b]{0.4\hsize}
  \begin{center}
   \rotatebox{-90}{%
   \includegraphics[scale=0.4]{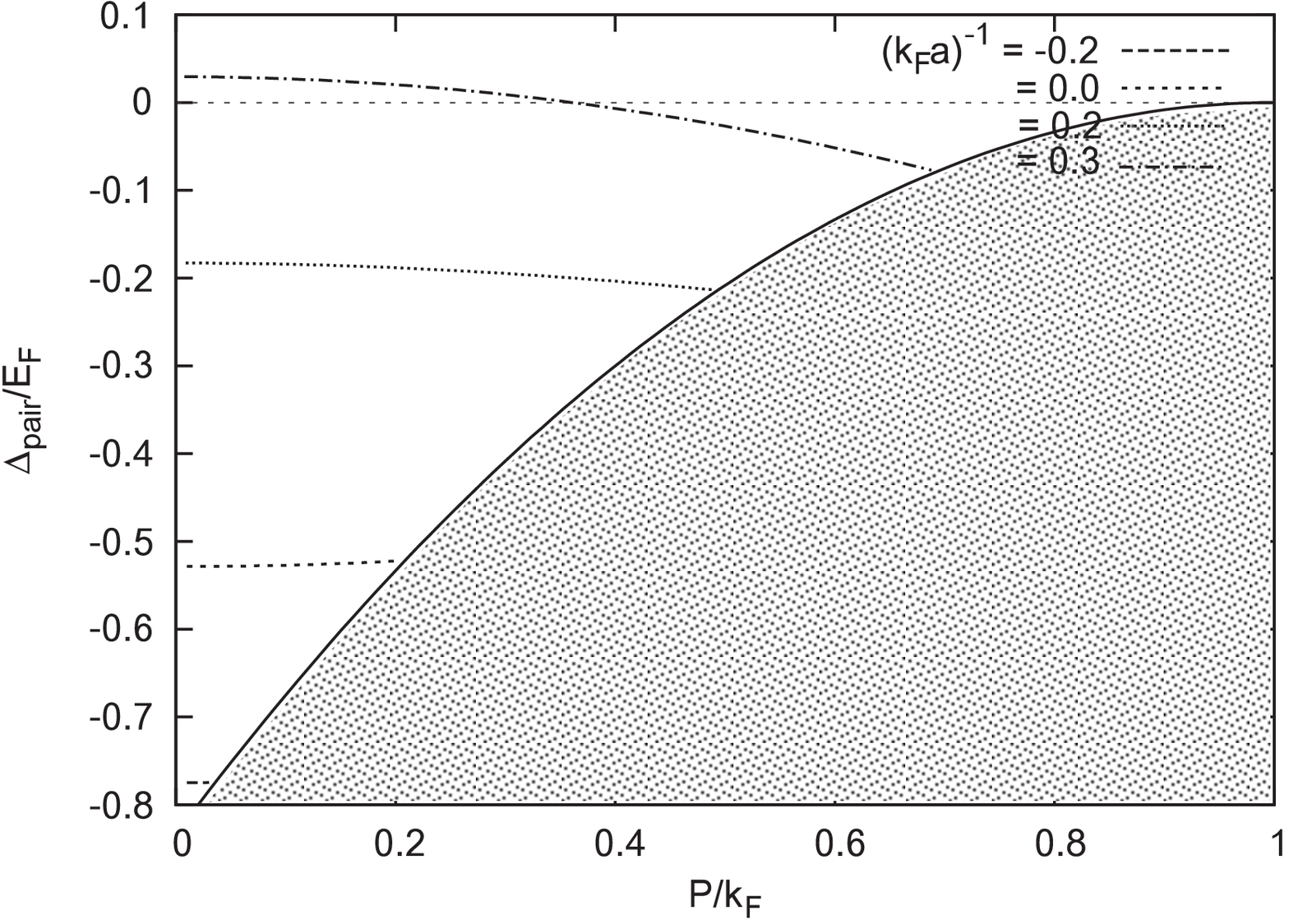}
   }
  \end{center}
 \end{minipage}
 \hspace{5mm}
 \begin{minipage}[b]{0.4\hsize}
  \begin{center}
   \rotatebox{-90}{%
   \includegraphics[scale=0.4]{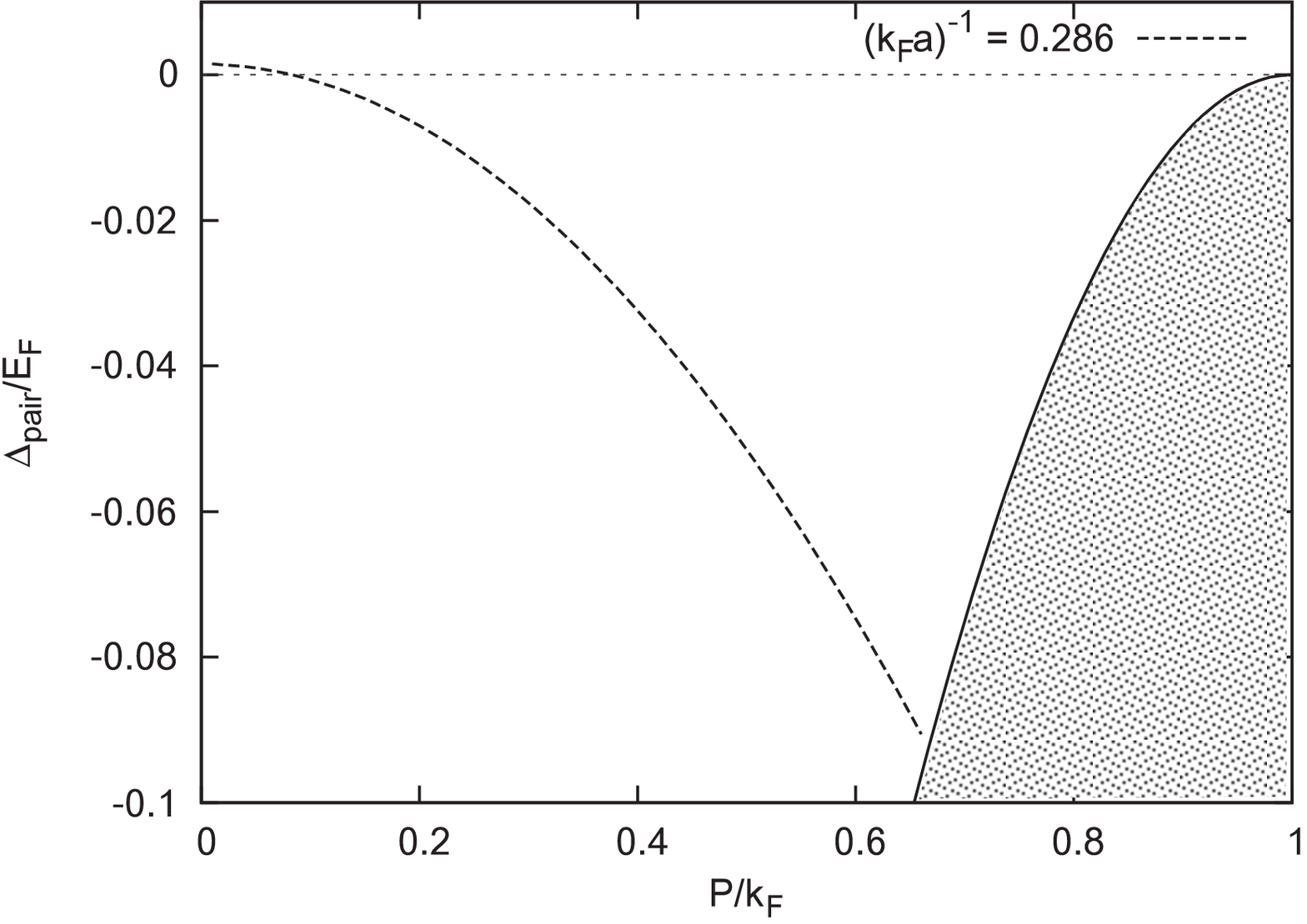}
   }
  \end{center}
 \end{minipage}
  \caption{Same as Fig.\ref{fig:delta_pair_m08} with $m_b/m_f = 1.2$.}
  \label{fig:delta_pair_m12}
\end{figure*}

We solved $P_0^c(|{\bf P}|)$ numerically. Here we include chemical potential $\mu_b$, 
and show $\Delta_{pair}(|{\bf P}|)$ in units of $E_F$ 
for  $\zeta=0.8,1.0,1.2$ in figs.\ref{fig:delta_pair_m08},
\ref{fig:delta_pair_m10} and \ref{fig:delta_pair_m12}. The number ratio is 
fixed at $N_b/N_f=1$.
Left panel of each figure shows results for several values of $1/k_Fa$.
One finds that each line extends up to a maximum value of $|{\bf P}|$, the energy of  which 
corresponds to $\epsilon_{\rm th}(|{\bf P}|)$ in Eq.(\ref{eq:eps_th}), 
i.e., the point where the pole hits the continuum and obtains a finite lifetime due to the 
non-vanishing imaginary part.
In the right panels of these figures, we show the dispersion curves at  the threshold 
values of $1/k_Fa$, where $\Delta_{pair}(|{\bf P}|)$ turns from negative to positive for 
the first time. 

We see from these figures that the effective binding increases as the interaction 
becomes more attractive, and eventually leads to a formation of the Bose-Fermi pair. 
The threshold value of $1/k_Fa$ is lower for a smaller value of $\zeta=m_b/m_f$ in 
agreement with the result for the total energy. We find also that 
$\Delta_{pair}$ is always negative in the unitarity limit, suggesting that the Bose-Fermi 
pair may not be formed in this limit. 

A peculiar feature seen from the figures is the non-monotonic dependence of 
$\Delta_{pair}$ against CM momentum $|{\bf P}|$, which is in contrast to the 
$P^2$ dependence of Eq.(\ref{eq:eps_coll}) in the simple picture. This is 
particularly apparent for small values of $\zeta=m_b/m_f$. Missing in the simple 
model is the $P$-dependence of the pair binding energy $E_{\rm rel}$ which should 
be present due, e.g., to the phase space available for the two-body scattering in medium. 
The medium effect becomes stronger for larger (positive) values of $(k_Fa)^{-1}$, 
especially for small $\zeta$ as seen from the figure. It may be mentioned in addition, 
that the results in the next subsection suggest a strong interplay of the collective and free branches 
for the Fermion dispersion relation.


\subsection{Fermion dispersion and level crossing}

Let us now investigate the pole structure of the single particle Green's function $G^f$ of (\ref{eq:Gfermi}).
The pole condition reads,
\begin{align}
	E_{\bf p} = \epsilon^f_{\bf p} + \Sigma^f(E_{\bf p}, {\bf p}) \label{eq:fermi_dispersion}.
\end{align}
In order to be consistent with our publication in \cite{schuck}, 
we here neglect the chemical potential $\mu_b$ in $\Gamma(P)$ of (\ref{eq:gammap}).
This means that we replace in $\Gamma(P)$ the Boson and Fermion propagators by totally free ones 
with the kinetic energies of boson and fermion and $\mu_b=0$ which is the uncorrelated value.

With the definition of $\Sigma^f$ in (\ref{eq:Gfermi}) with $\mu_b=0$ 
and the expression $\Gamma$ of (\ref{eq:gammap}), 
we easily can investigate the poles of $G^f$ for various system parameters. 
We first consider cases, like in \cite{schuck}, where $N_b \ll N_f$.

We expect two branches: one which corresponds in week coupling to $E_{\bf p}\sim \epsilon^f_{\bf p}$ 
and one which corresponds to the collective branch, 
i.e. the pole contained in $\Gamma(p)$.

\begin{figure*}[htb]
  \begin{center}
    \begin{tabular}{cc}
      \resizebox{75mm}{!}{\rotatebox{-90}{\includegraphics{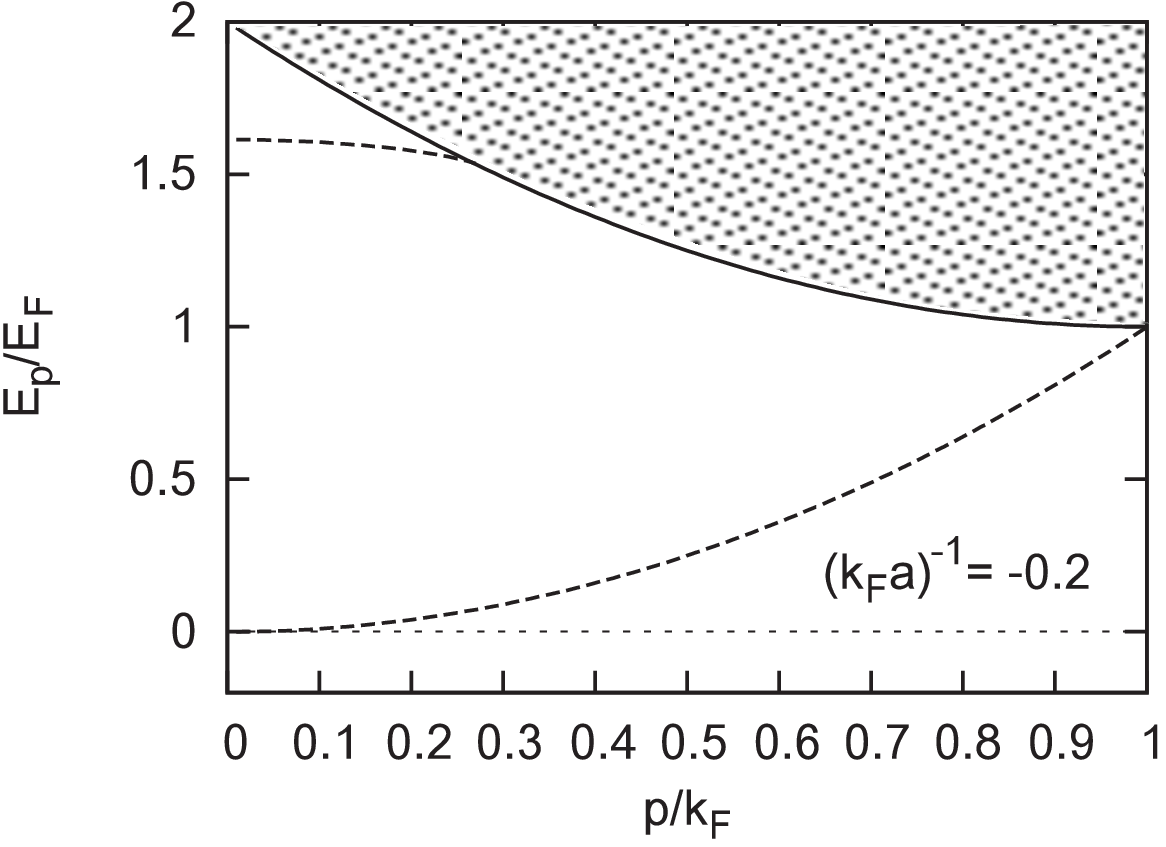}}} & 
		\resizebox{75mm}{!}{\rotatebox{-90}{\includegraphics{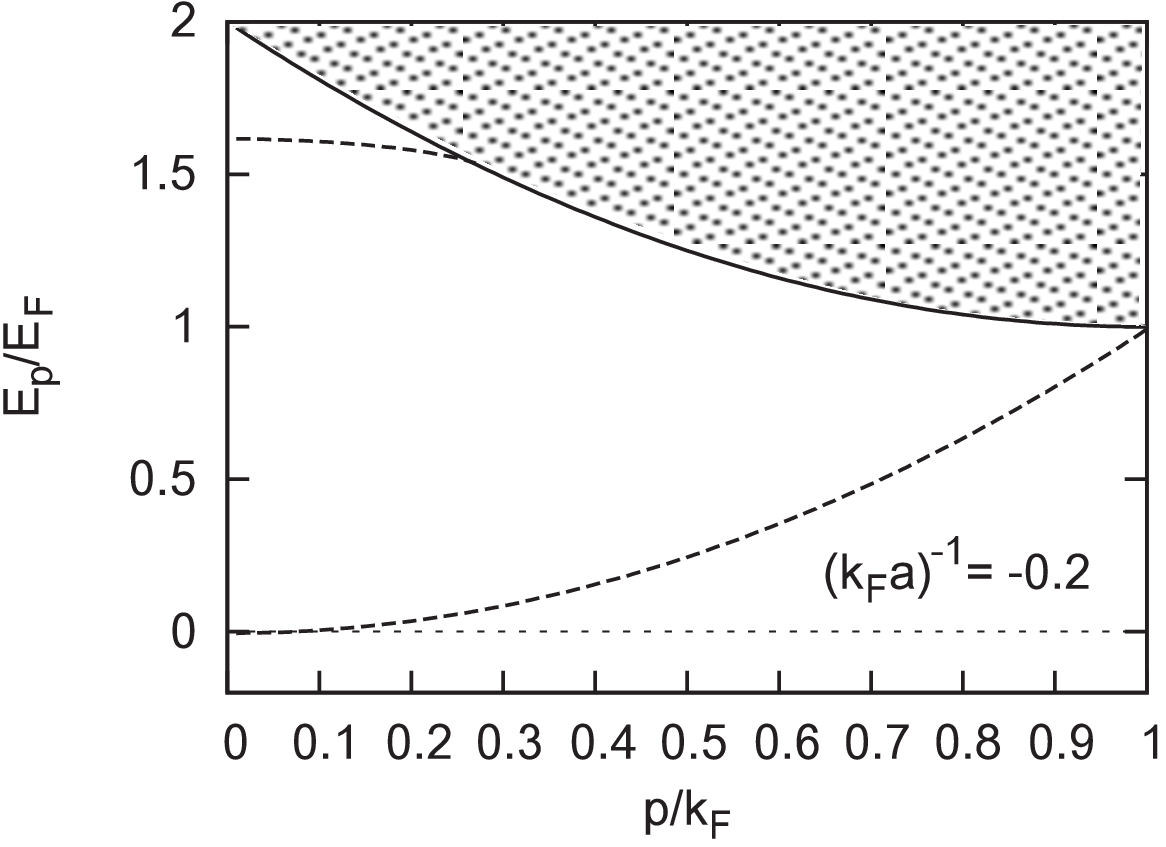}}} \\
      \resizebox{75mm}{!}{\rotatebox{-90}{\includegraphics{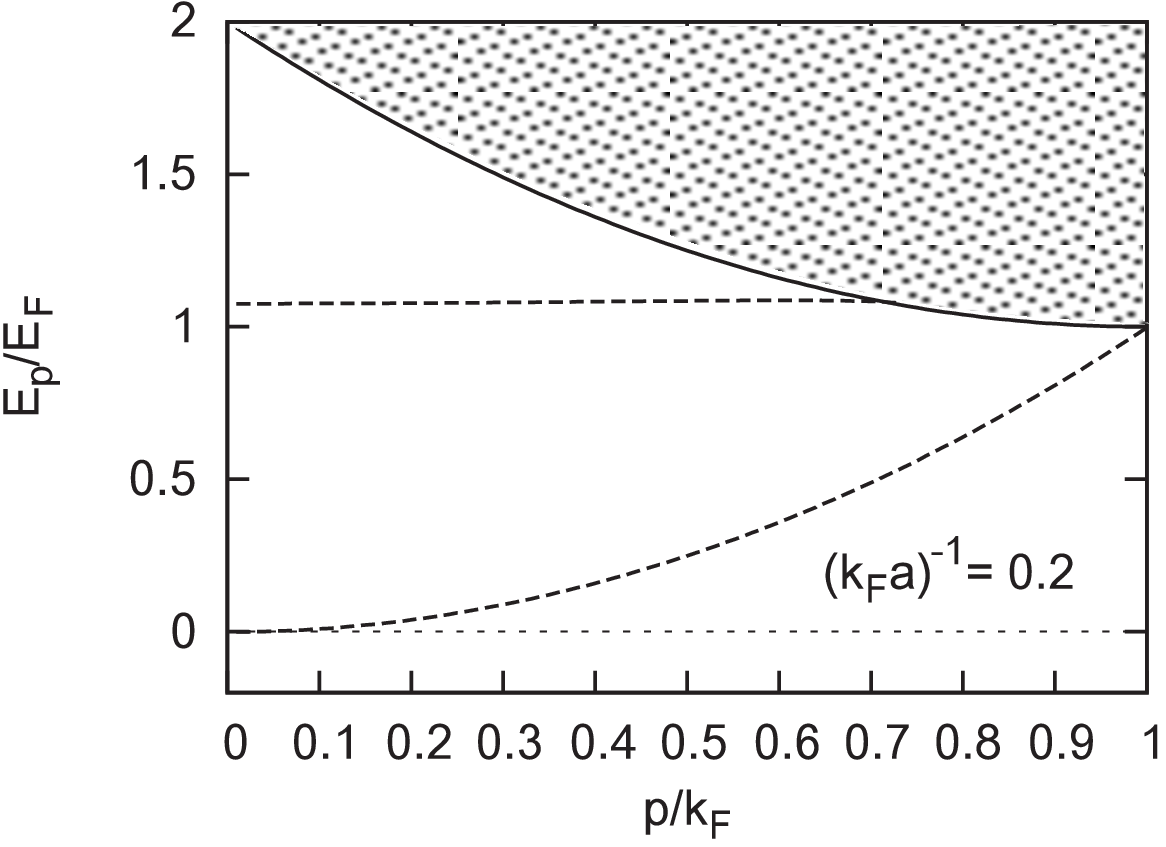}}} & 
		\resizebox{75mm}{!}{\rotatebox{-90}{\includegraphics{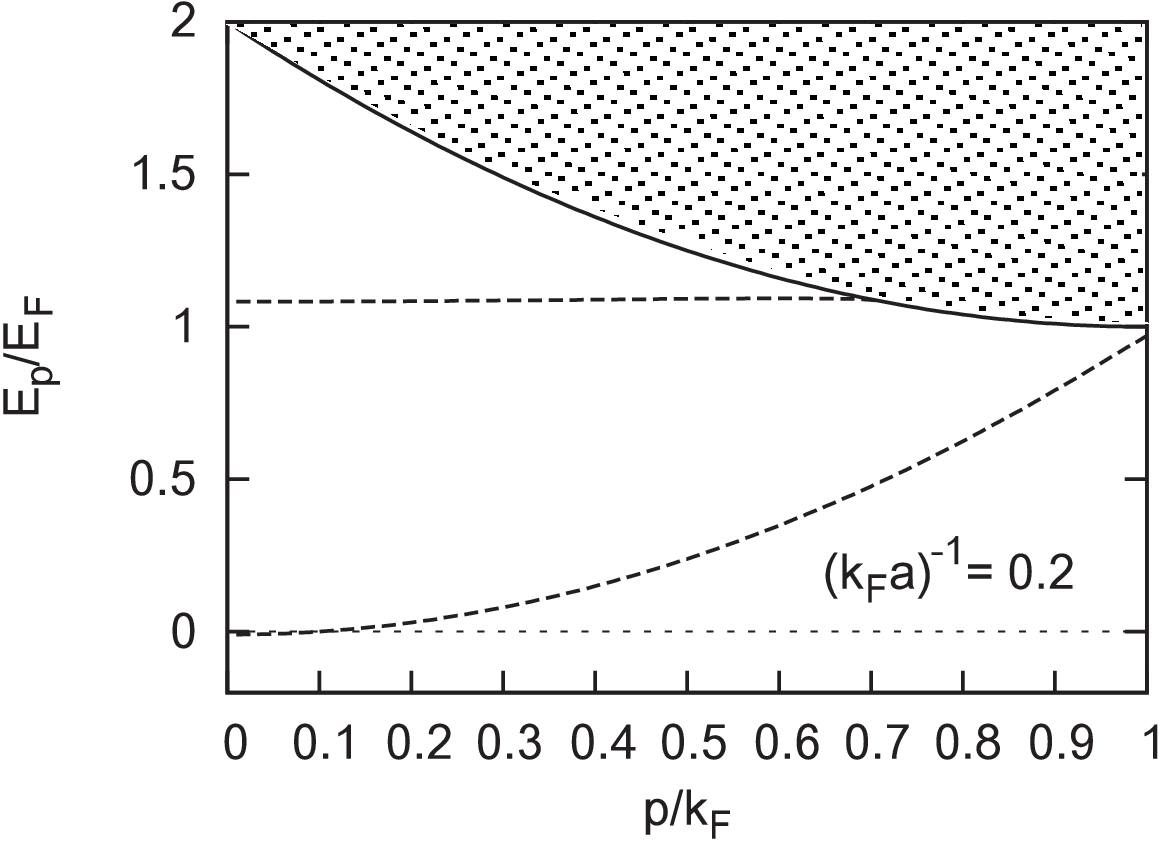}}} \\
      \resizebox{75mm}{!}{\rotatebox{-90}{\includegraphics{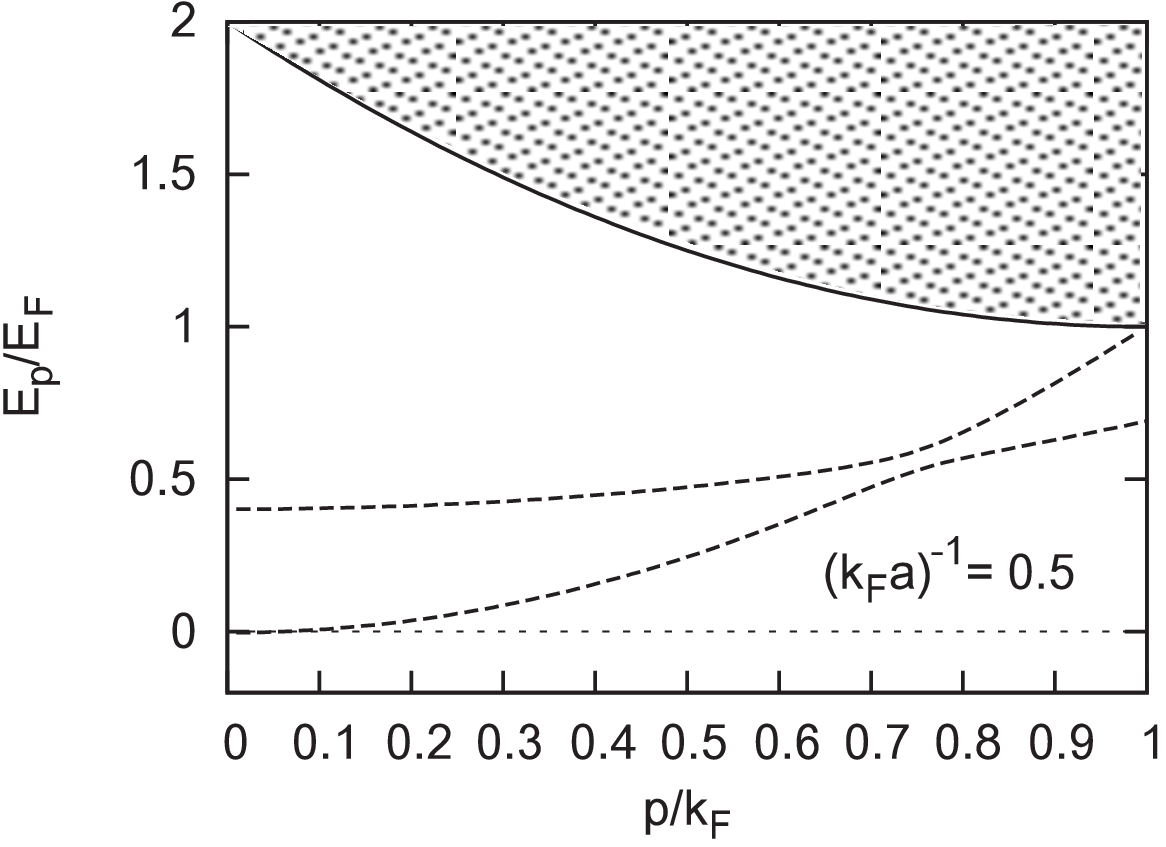}}} & 
		\resizebox{75mm}{!}{\rotatebox{-90}{\includegraphics{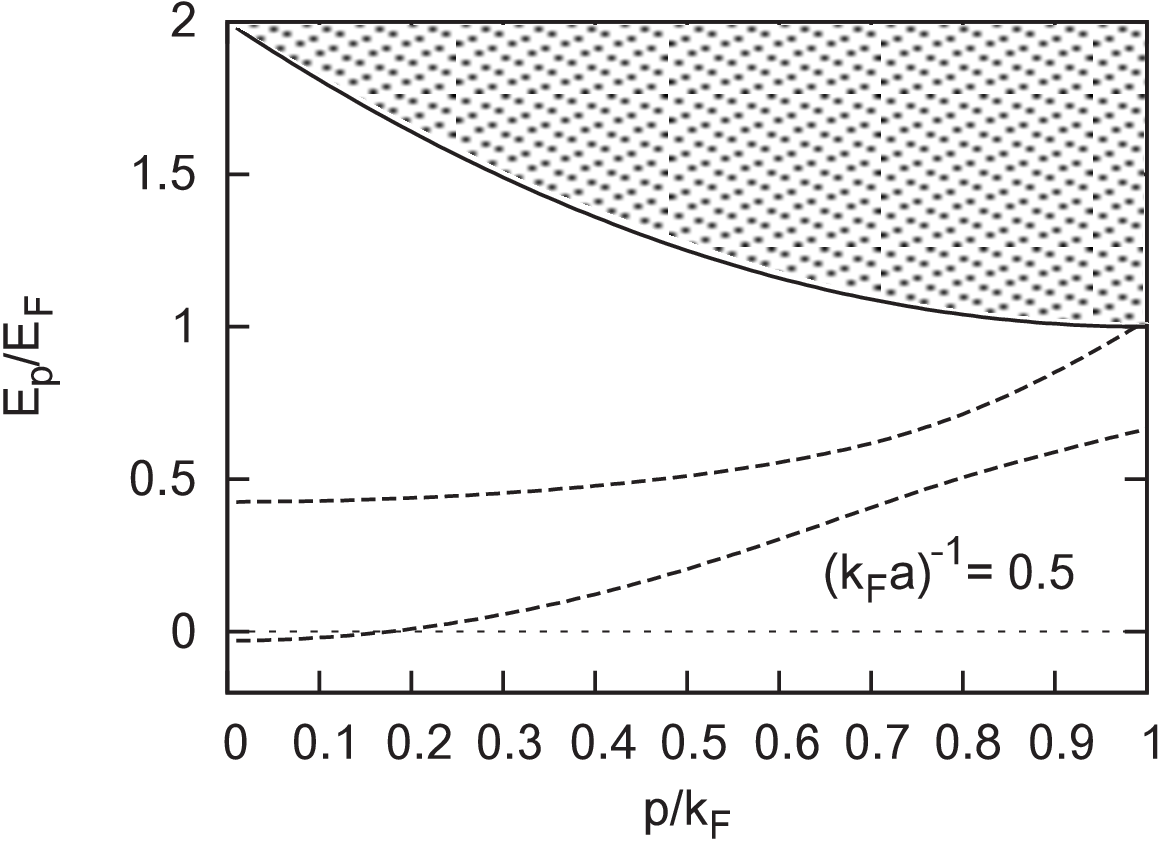} }} \\
    \end{tabular}
    \caption{Fermion dispersion curves. Left row corresponds $N_b/N_f=0.001$ and right row corresponds
			$N_b/N_f=0.01$. From top line to bottom,$(k_Fa)^{-1}$ are $-0.2$,$0.2$ and $0.5$.
			$m_b/m_f$ is fixed to $1.0$.}
    \label{fig:fermi_disp_two}
  \end{center}
\end{figure*}
In fig.\ref{fig:fermi_disp_two}, we show on the left panel the case of $N_b/N_f=1/1000$ and on the 
right panel the case of $N_b/N_f=1/100$. 
From top to botom, we have $(k_Fa)^{-1}=-0.2, 0.2, 0.5$.
Here mass ratio $m_b/m_f=1$. 
The dotted area is the region where the imaginary part of $\Sigma^f$ is different from zero. 
As in our ealier work \cite{schuck}, 
we see an avoided crossing of the two branches at some finite value of $p/k_F$. 
We also see that the interaction between the two branches becomes stronger as $N_b/N_f$ increases.

\begin{figure*}[htb]
 \begin{minipage}[b]{0.4\hsize}
  \begin{center}
   \rotatebox{-90}{%
   \includegraphics[scale=0.55]{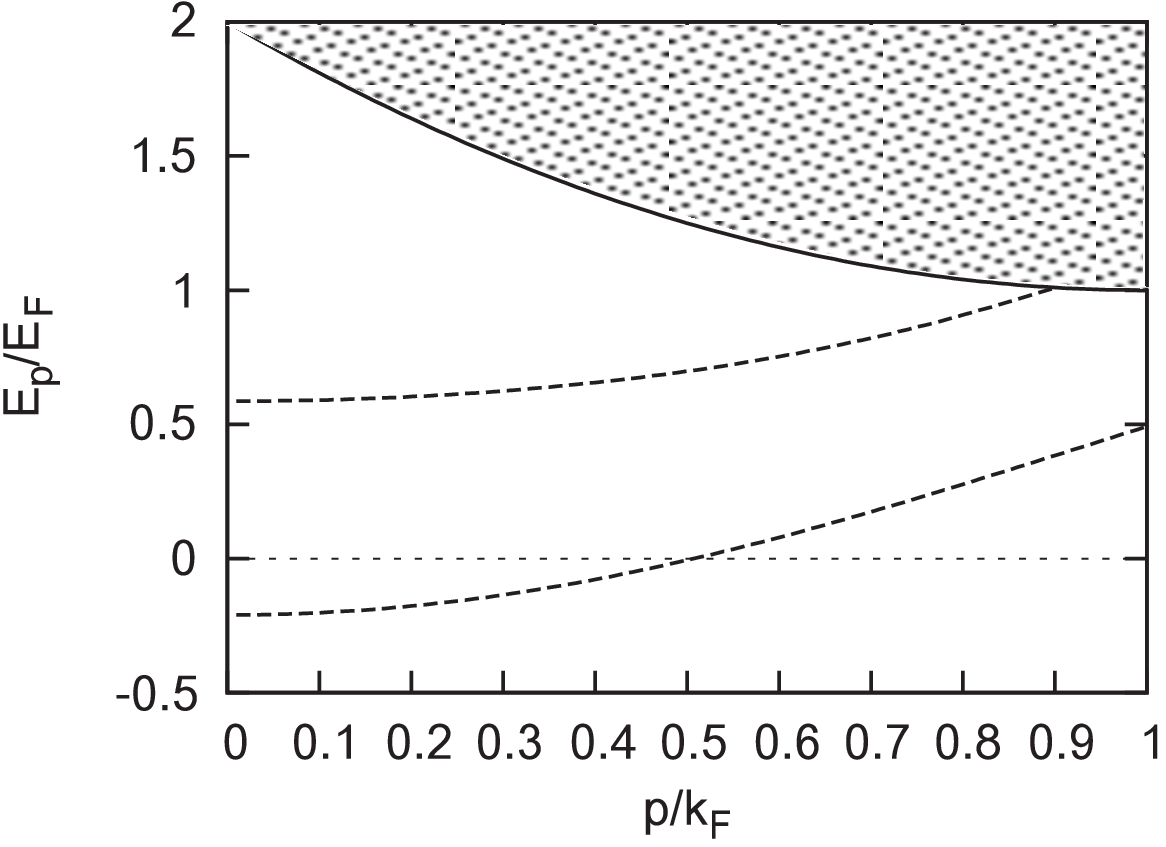}
   }
  \end{center}
 \end{minipage}
 \hspace{5mm}
 \begin{minipage}[b]{0.4\hsize}
  \begin{center}
   \rotatebox{-90}{%
   \includegraphics[scale=0.55]{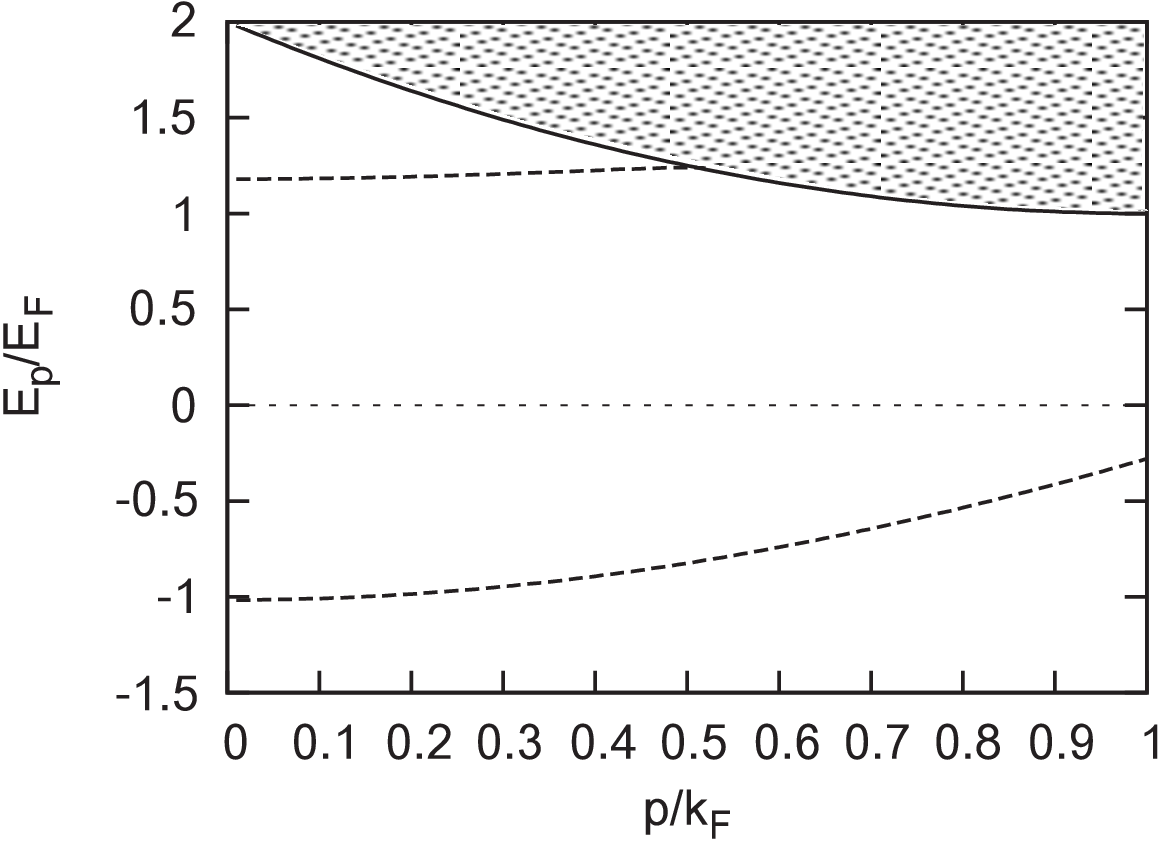}
   }
  \end{center}
 \end{minipage}
  \caption{Fermion dispersion curves. Left panel corresponds $N_b/N_f=0.1$ and right one corresponds $N_b/N_f=1.0$.
		$(k_Fa)^{-1}$ are fixed to $0.5$. And $m_b/m_f=1.0$.}
  \label{fig:fdisp_a05}
\end{figure*}
In fig.\ref{fig:fdisp_a05}, we show the case $(k_fa)^{-1}=0.5$ and 
$N_b/N_f=1/10$ for left panel and $N_b/N_f=1$ for right panel. 
We see that no crossing feature is visible any longer and the two branches probably become completely hybridised. 
In a future publication we intend to investigate in a systematic way the nature of the two branches 
as a function of the system parameters.

\section{Summary, Discussion and Conclusion}

We studied in this paper the static properties of the Bose-Fermi mixture in the 
lowest order of the hole-line expansion, 
i.e. in the T-matrix approximation. 
The interaction parameter is expressed 
in terms of the scattering length up to infinite order using the renormalization 
procedure of ref.\cite{randeria}, so as to allow for the calculation around the 
unitarity limit. 
The T-matrix approach is a common approximation 
often used in the past in Fermi and Bose systems \cite{fetter}. 
It has, however, not been investigated very much in Bose-Fermi mixtures and, therefore, 
one has not much experience about its quality in that situation. 
Recently appeared, however, a study for a one dimensional system \cite{pollet}, 
from where exact Quantum Monte Carlo results are available for comparison. 
In ref \cite{xavier} it is shown that the T-matrix approach yields quite reasonable results also in the Bose-Fermi case, 
even in a one dimensional case which is probably the worst situation possible.

With this background in mind, 
we first studied energy and pressure as functions of the inverse scattering 
length for several choices of the mass ratio $\zeta=m_b/m_f$ and the number ratio 
$N_b/N_f$. The energy of the system becomes strongly attractive as the inverse 
scattering length changes sign from negative to positive, i.e. around the unitary limit. 
As one increases the number of bosons with respect to the fermions, 
arrives a point where the pressure becomes negative, 
i.e. the system becomes unstable (collapse). 
The effect is stronger for small values of $m_b/m_f$. 
This is not in contradiction with experiments \cite{BFcollapse}.

Next we studied the possibility of stable BF-pairs, as in \cite{schuck}. 
We, indeed, also found in the present model that even for infinitesimal BF-attraction 
a stable BF-mode appears, reminiscent of the Cooper pole in a two component Fermi gas, 
since in both cases its origin stems from the presence of a sharp Fermi surface. 
However, in the BF case the BF-pair is a composite fermion, 
whereas in the original Cooper problem, one has a composite fermion pair, 
i.e. a boson-like cluster. In the latter case, 
many pairs can well be treated by the usual BCS formalism. 
On the other hand , the case of many BF-pairs still has to be worked out. This shall be done in future work.
But we here studied the BF pair formation 
by observing its binding energy measured from the last filled free BF pair. 
For some finite values of the attractive interaction, 
there occurs the formation of stable BF pairs.
For $\zeta\geq 1$ the pair shows a standard dispersion of a quasiparticle, while for $\zeta=0.8$ the 
pair with finite center-of-mass momenta feel stronger attraction. This effect 
is not clearly seen in the energy or pressure, where the singular effect may have 
been averaged out.

Up to this point we assumed an ideal case where the Bose-Fermi interaction is 
dominant, while the Bose-Bose interaction was neglected. To compare with a real 
atomic gas system, the Bose-Bose interaction cannot be discarded 
even when the Bose-Fermi interaction is enhanced, e.g., through Feshbach resonances. 
We checked the effect of the Bose-Bose interaction on the energy of the 
system up to the first order in $N_b^{1/3}a_{bb}$.  We took the BoseBose 
scattering length $a_{bb}$ in the range $-0.3\leq k_F a_{bb}\leq 0.3$, and the 
number ratio $N_b/N_f=$2.0, 1.0 and 0.5, and repeated the calculation of 
the $\beta$ parameter and the pressure \cite{watanabe}.
The calculation shows that the effect of the Bose-Bose interaction is small within 
the adopted parameter values: The $\beta$ value at the unitarity limit, for instance, 
deviates less than 10\% for 
$m_b=m_f$ and $N_b=N_f$, and does not change conclusions obtained at $a_{bb}=0$.

In summary, we suggest that the energy gain in the Bose-Fermi mixture at positive 
values of $a_{bf}$ is related to the formation of the resonant Bose-Fermi pairs, and 
that the center-of-mass momenta of the pairs are dependent on the ratio $m_b/m_f$ 
due to the statistics of the two kinds of the particles. 

\begin{acknowledgments}
We thank X. Barillier-Pertuisel, K. Suziki, T. Nishimura, T. Maruyama and H. Yabu for useful discussions. 
We also thank J. Dukelsky and S. Pittel for their general interest 
and contributions to the subject of BF correlations.
\end{acknowledgments}

\appendix*
\section{Expression for the Scattering Amplitude}

Let us calculate the $I(P_0,|{\bf P}|)$ (\ref{eq:gammap}) of the scattering amplitude 
$\Gamma(P)$ 
\begin{widetext}
\begin{equation}
        I(P_0, |{\bf P}|)) = \int \frac{d^3{\bf k}}{\left( 2\pi \right)^3}
         \left\{ \frac{ \theta \left( \left| \tilde{\bf P}_f + {\bf k} \right| - k_F \right) }
         {P_0-\epsilon^f_{{\bf \tilde{P}}_f + {\bf k}} -\epsilon^b_{{\bf \tilde{P}}_b -{\bf k}} + \mu_b +i \eta}
         + \frac{1}{\epsilon^f_{\bf k} + \epsilon^b_{\bf k} } \right\}.
\end{equation}
\end{widetext}
As we are interested in the real part of the pole of $\Gamma(P)$, we hereafter omit 
$i\eta$ in the denominator. Each term in the integrand shows an ultraviolet divergence, and 
we formally introduce a cutoff $\Lambda$ which will be taken to infinity later.
\begin{widetext}
\begin{equation}
        I_1(P_0, |{\bf P}|) = \frac{1}{\left( 2\pi \right)^2} 
             \lim_{\Lambda \to \infty} \int^{\Lambda}_{k_F} dk \, 
              \left[ \left( - \frac{m_b}{|{\bf P}|} \right) k \ln  \left| 
         \frac{ \frac{k^2}{2\nu} + \frac{|{\bf P}|k}{m_b} -P_0 + \frac{{\bf P}^2}{2m_b}-\mu_b }
              { \frac{k^2}{2\nu} - \frac{|{\bf P}|k}{m_b} -P_0 + \frac{{\bf P}^2}{2m_b}-\mu_b } \right| 
          +2k^2 \frac{1}{\frac{k^2}{2\nu}} \right] .
         \end{equation}
The divergent term at $\Lambda\rightarrow\infty$ coming from the first term in the integrand 
is cancelled out by the second term, and we obtain the finite quantit 
\begin{equation}
        I(P_0. |{\bf P}|) = \frac{1}{\left( 2\pi \right)^2} \left( \frac{m_b}{2|{\bf P}|} \right)
           \Bigg[  \left\{ k_F^2 - \left( \frac{\nu |{\bf P}|}{m_b} \right)^2 - A \right\} 
             \ln{ \left| \frac{ \left( k_F + \frac{\nu |{\bf P}|}{m_b} \right)^2 - A }
           { \left( k_F - \frac{\nu |{\bf P}|}{m_b} \right)^2 - A } \right| } 
               +  \frac{4\nu |{\bf P}|}{m_b}k_F   - \frac{4\nu |{\bf P}|}{m_b} A F(A)  \Bigg] ,
        \end{equation}
where 
\begin{equation}
      F(A) =  \begin{cases}
          \frac{-1}{2\sqrt{A}} \ln{ \left| \frac{ \left( k_F + \frac{\nu |{\bf P}|}{m_b} - \sqrt{A} \right)
              \left( k_F - \frac{\nu |{\bf P}|}{m_b} - \sqrt{A} \right) }
              {  \left( k_F + \frac{\nu |{\bf P}|}{m_b} + \sqrt{A} \right) 
                 \left( k_F - \frac{\nu |{\bf P}|}{m_b} + \sqrt{A} \right) }   \right| } & (A > 0) \\
          \frac{2k_F}{ k_F^2-\left( \frac{\nu |{\bf P}|}{m_b} \right)^2 } & (A = 0) \\
           \frac{\pi}{\sqrt{-A}} - \frac{1}{\sqrt{-A}} 
                    \arctan{ \left( \frac{k_F+\frac{\nu |{\bf P}|}{m_b}}{\sqrt{-A}} \right) }
           - \frac{1}{\sqrt{-A}} \arctan{ \left( \frac{k_F-\frac{\nu |{\bf P}|}{m_b}}{\sqrt{-A}} \right) } 
                                                        & (A < 0) 
          \end{cases}
          \end{equation}
\end{widetext}
with
\begin{equation}
        A = \left( \frac{\nu |{\bf P}|}{m_b} \right)^2 - 2\nu
                                 \left( -P_0 + \frac{{\bf P}^2}{2m_b} - \mu_b \right).
        \end{equation}
The final expression for $\Gamma(P)$ is given in terms of $I(P)$ by
\begin{equation}
       \Gamma(P)=\frac{2\pi a}{\nu} \Bigg[ 1- \frac{2\pi a}{\nu}I(P_0, |{\bf P}|) \Bigg]^{-1}.
       \end{equation}




\end{document}